\documentclass{aa}
\usepackage{graphicx}
\begin{document}

  \title{Radio emission from shell-type supernova remnants}

   % \subtitle{}

   \author{A.I.Asvarov}
         
   \offprints{A.I.Asvarov}

   \institute{Institute of Physics, Azerbaijan National Academy of Sciences, BAKU AZ1143, Azerbaijan\\
              \email{asvarov@physics.ab.az}}
          
   \date{Received month day, year; accepted month day, year}
   
   \authorrunning{Asvarov A.I.}
   \titlerunning{Radio emission from SNRs}

\abstract { 
 The evolution of the radio emission of shell-type Supernova remnants (SNRs) is modeled within the framework of the simple and commonly used assumptions that the mechanism of diffusive shock acceleration (DSA) is responsible for generating radio emitting electrons and that the magnetic field is the typical interstellar field compressed at the shock. It is considered that electrons are injected into the mechanism in test-particle regime directly from the high energy tail of the downstream Maxwellian distribution function. The model can be applied to most of the observed SNRs because the  majority of detected SNRs are shell-types and have a more or less spherical shape and are sources of nonthermal radio emission. It is shown that the model successfully explains the many averaged observational properties of evolved shell-type SNRs. In particular, the radio surface brightness ($\Sigma$) evolves with diameter as $\sim D^{-(0.3 \div 0.5)}$, while the bounding shock is strong (Mach number is ${\mathcal M} \geq10$), followed by steep decrease (steeper than $\sim D^{-4.5}$) for ${\cal M} <10$. 
Such evolution of the surface brightness with diameter and its strong dependence on the environmental parameters strongly reduce the usefulness of  $\Sigma - D$ relations as a tool for determining the distances to SNRs. The model predicts no radio emission from SNRs in the late radiative stage of evolution and the existence of radio-quiet but relatively active SNRs is possible. 
Our model easily explains very large-diameter radio sources such as the Galactic Loops and the candidates for Hypernova radio remnants.
The model predicts that most of the observed SNRs with $ \Sigma _{{\rm 1GHz}} \la \,\,10^{ - 20} \,\,{\rm W}\,{\rm m}^{{\rm  - 2}} \,{\rm sr}^{{\rm  - 1}} \,{\rm Hz}^{{\rm  - 1}}$ are located in a tenuous phase of the ISM.  
The model also predicts the existence of a population of $150-250$~pc SNRs with $ \Sigma _{{\rm 1GHz}}  \la \,\,10^{ - 22} \,\,{\rm W}\,{\rm m}^{{\rm  - 2}} \,{\rm sr}^{{\rm  - 1}} \,{\rm Hz}^{{\rm  - 1}}$ if the kinetic energy of the explosion is $\sim 10^{51}$~erg. 
From the comparison of the model results with the statistics of evolved shell-type SNRs, we were able to estimate the fraction of electrons accelerated from the thermal pool in the range $(3\div 11) \times 10^{ - 4}$. If acceleration takes place directly from the high energy tail of the downstream Maxwellian distribution function, then the corresponding injection momentum is estimated as $p_{\rm inj}\sim (2.7-3)\cdot p_{\rm th}$

   } 
   \maketitle
%________________________________________________________________

\section{Introduction}

The radio synchrotron radiation is a prominent output of SNRs and is normally used to identify them among other radio sources. 
Although SNRs are studied at almost every wavelength ranging from gamma-rays through low-frequency radio waves, the radio observations still remain the most effective means of obtaining important information on these objects; in fact, 
most of the Galactic SNRs have been discovered by their radio emission. Naturally, much richer observational information on the SNRs is available in the radio range; in the last version ( January 2004) of the Green (\cite{green}) catalogue 231 remnants are listed, a number that is continuously growing (e.g., Brogan et al. \cite{brogan}). For most recent review of the radio SNRs, see Reich (\cite{reich}).

 The question of the origin of the high-energy electrons and magnetic fields responsible for the synchrotron radio emission from the SNR shells has still not been fully solved. Since the pioneering works on particle acceleration at collisionless shocks by Krymskii (\cite{kr77}), Bell (\cite{b78a}), and Blandford \& Ostriker (\cite{bo78}), the DSA  mechanism have been regarded as the most suitable mechanism responsible for the origin of radio-emitting electrons in  shell-like SNRs. 
Although we directly observe only electrons, the mechanism of DSA has theoretically more successful application to the protons and heavy ions. The effectiveness of the acceleration for ions may be very high in certain parameters regimes (see the review by Malkov \& Drury \cite{maldr}), and this mechanism is believed to be the source of Galactic cosmic rays with energies up to $10^{15} \,{\rm eV}$.     
 In the case of electrons,  not only is the electron acceleration theoretically problematical but the mechanism for electron heating in collisionless shocks is also poorly understood. At the same time, synchrotronically radio-emitting relativistic electrons in shell-type SNRs with spectra coincident with the prediction of the mechanism can be considered as the only direct evidence of a DSA action. The recently observed X-ray synchrotron emission from several SNRs has strengthened this evidence (see Ballet \cite {Ballet05} for a recent review).

There are a number of models describing the evolution of the radio emission of young SNRs (e.g., Chevalier \cite {ch1982}; Gull \cite{gul73}; Dickel et al. \cite{dic89}) and of the SNRs in cloudy environments (e.g., Chevalier \cite{ch1999}; Bykov et al. \cite{byk2000}).  
 The most popular model for an interpretation of the radio emission from evolved and old SNR is the mechanism of van der Laan (\cite{laan}) and its modifications. According to this model the magnetic fields and the interstellar cosmic ray electrons compressed behind a cooling shock front are responsible for the continuum synchrotron radio emission of SNRs, so this mechanism is applicable to the remnants in the radiative phase of evolutions.
 In spite of the fact that the general theory of SNR  predicts the inevitable onset of a radiative phase in the course of SNR evolution (e.g., Chevalier \cite{ch1974}), observationally  very few candidates for radiative SNRs are detected.
 %Although according to the general theory of shell-type SNR the radiative phase in the course of SNR evolution is inevitable (e.g., Chevalier \cite{ch1974}), observationally  very few candidates for radiative SNRs are detected.
  The set of SNRs, which have long been considered as highly evolved remnants in the radiative phase, include IC 443,  W44, CTB 80, S147, MSH 11-61A, OA 184, etc. However, the intensive optical emission observed from these SNRs in most cases can be interpreted by the interaction of a non-radiative adiabatic shock wave with the dense interstellar clouds. The co-existence of intensive optical emission and high-temperature plasma, which gives the thermal X-rays, is considered as an argument in favor of such interpretation. Indeed, in several SNRs there is direct evidence of slow radiative shocks, which are driven into ambient clouds by a fast nonradiative blast wave (McKee \& Cowie \cite{mckeecow75}). For instance, in Seta et al. (\cite{seta98}) clear evidence of the interaction of SNRs W44 and IC 443 with the molecular clouds is presented. In the Cygnus Loop SNR, the bright optical filaments are attributed to shocked clouds (e.g., Hester \& Cox  \cite{HCox}; Patnaude et al. \cite {patnaud}). Shocked clouds have also been observed in Vela SNR (e.g., Bocchino et al. \cite{bacchino99}), so there are very few candidates for SNRs that could be considered as real radiative remnants in accordance with the standard theory. The S147 SNR (Reich \cite{reich}) can be considered as a representative of this tiny group of SNRs. 

 The lack of radiative SNRs in our galaxy can be explained in several ways.
First, there are inevitable selection effects, mainly due to the irregular high-intensity background emission, which works against the detection of extended SNRs evolving in the rarefied environs. Second, as shown (Blondin et al. \cite{blon98}; Bertschinger \cite{bert86}; Vishniac \cite{vish83}; Falle \cite{fale}), the shells of SNRs at the beginning of the radiative stage become highly unstable leading to fragmentation of the dense shell,  thereby making the action of the  Laan's mechanism become inefficient. And, finally, if the density of the medium in which the SNR is located is low enough, it is possible for the SNR to finish its life by merging with the ISM before cooling becomes important. In addition, it is important to take into account the effect of accelerated particles, trapped in the  matter behind the shock front, which may impede the processes of catastrophic cooling.  %
The problem of the visibility of old radiative SNRs in the {\sc Hi} 21-cm emission line in our Galaxy was considered in the recent work by Koo \& Kang (\cite{koo}), who also noted a considerable deficit of radiative SNRs in our Galaxy.

As a result, the van der Laan mechanism encounters a number of difficulties in interpreting the radio emission from the observed large-diameter SNRs. However, under some specific conditions, the modification of this mechanism proposed by Blandford \& Cowie (\cite{bcow82}) (see also Bychkov \cite{bychkov}) can be effective in generating  radio emission from old and large SNRs. Although these models also include the mechanism of DSA,  the specific cloudy ambient interstellar medium is required to be effective in generating radio emission; besides, the physics of cloud - shock interaction is far from being understood in detail.

Though the ISM generally has an inhomogeneous nature, many extended and old SNRs, as a rule, demonstrate nearly circular shells in radio and x-ray, which implies that they have been expanding into a relatively homogeneous region of the ISM. In practice, the circular shape is one of the observational properties of SNR by which they are identified as SNR.  In such homogeneous regions, it is possible, however, that such small-scale clouds  exist (Nagashima et al. \cite {Nagashima06}), as required by the model of Blandford \& Cowie (\cite{bcow82}). Two cases are possible. If small-scale clouds have high spatial concentration, the evolution of SNR occurs according to the scheme considered by Cowie et al. (\cite{CMO81}), which will result in the morphology being completely different from the standard Sedov solution (see also Chieze \& Lazareff \cite{chilaz81}). Indeed, a small fraction of SNRs in X-rays show the morphology consistent with such a possibility. According to Rho \& Petre (\cite{rho}), up to 25\% of all X-ray-detected Galactic SNRs show this ``mixed-morphology'': i.e., center-filled X-ray and shell-like radio morphology. Moreover, in such a medium the SNR cannot extend up to the large diameters, which allows us to exclude this case from consideration. In the other case of a low  concentration of the interstellar cloudlets,  their contribution to the integral  radio emission will be minor, and the remnant will have a more or less regular shell morphology. In this case the mechanism responsible for generating the radio-emitting relativistic electrons works at the outer shock front that regularly bounds the SNR.
In reality, we do not exclude the possibility that the van der Laan mechanism  (\cite{laan}) will work locally in some parts of the large diameter remnant; in other words, different stages can occur simultaneously in different locations within a single remnant. Very probably this is the case in above-mentioned, optically bright galactic SNRs.

Recently Berezhko \& V\"olk (\cite{berezhko}) developed a time-dependent model of radioemission from shell-type SNRs, taking into account the   non-linear shock acceleration effect and the generation of the magnetic field in the shock vicinity. This model is very good for young SNRs, but in case of evolved and old remnants, the application of both of these effects can be questioned.% (???) but strongly complicates the model.
Therefore, it is important to construct a model that is based on the very general results of the DSA theory, and that resorts to minimal non-standard assumptions.
In this paper we suggest the model describing the evolution of radio emission of SNRs, which is based on the very common assumptions that the radio-emitting electrons are accelerated diffusively at the shock front from the thermal pool in the test-particle regime. %
The aim of the work is not to model a specific SNR in detail but to illustrate the general behavior of the radio properties of SNRs (surface brightness, spectral index, etc.) with their evolution, to test the capability of DSA in reproducing the general statistics of the shell-type SNRs, and to constrain the injection problem as far as possible.

The set of main assumptions are the following:

  -- We consider spherical adiabatic SNR evolving in the homogeneous ISM. To follow the evolution of the remnant up to the maximum sizes,  we used an analytical approximation of Cox \& Andersen(\cite{coxa82}), which describes the evolution of an adiabatic, spherical blast wave in a homogeneous ambient medium of finite pressure. At early times while the shock wave is strong, this approximation coincides well with the zero-pressure self-similar Sedov solution. In applications of the model to real SNRs, the pressure of the interstellar magnetic field is included.

  -- Electrons are accelerated diffusively at the main shock front, and their injection takes place from the downstream thermal distribution function.  The threshold energy of injection  is proportional to the temperature of the downstream plasma, and as a consequence of this assumption, the injection of new particles in the process of acceleration stops when radiative cooling of the shocked gas first starts. The test particle approximation is used, all the nonlinear effects are neglected, and electron-ion temperature equilibrium is presumed. It is also assumed that accelerated electrons only lose their energy adiabatically and there is no diffusion. %In applications of the model to real SNRs.

  -- The magnetic field is the typical interstellar field compressed at the shock wave up to four times; i.e. there is no additional amplification of the interstellar magnetic field in the shell of the SNR. It is also assumed that it is fully chaotic and is frozen-in to the plasma inside the remnant.

The paper is organized as follows. In Sect. 2 the radially-dependent distribution function of accelerated electrons is derived, which was  then used to obtain the basic formulas describing the radio properties of SNR. The list of main predictions of the model is presented at the end of the section. To apply our model to real SNRs  the initial conditions and parameters are specified in  Sect. 3.  Further, the empirical $\Sigma - D$ relation is constructed to compare the predictions of the model with observational data . With the help of this relation, the values of the injection parameter is estimated. The final section contains our main conclusions.

\section{Model}

\subsection {Distribution function of accelerated electrons}
	
		In the test particle approximation, the action of DSA in planar, steady shock front leads to an isotropic momentum distribution function for the accelerated particles of the form
		
\begin{equation}
f_{\rm a}  = Ap^{ - q} \left[ {H(p - p_{{\rm inj}} ) - H(p - p_{{\rm max}} )} \right]\label{eq1}
\end{equation}
where $A$ is the normalization constant, $H$  the Heaviside step function, $p_{\rm inj}$   the injection momentum, and $p_{\rm max}$ the maximum momentum of electrons. A very important property of this spectrum is that the spectral index $q$, defined as 

\begin{equation}
q = \frac{{3x_{\rm s} }}{{x_{\rm s}  - 1}},
\label{eq2}
\end{equation}
depends only on the shock compression ratio $x_{\rm s} = \rho /\rho _0$, where $\rho_0$ and $\rho$ are mass densities for the up- and downstream states, respectively. As a parameter describing the evolution of SNR in the following  we mainly use the Mach number defined as the ratio of shock velocity  ${\rm v}_{\rm s}$ to the sound speed (correctly, to the maximum signal velocity ahead of the shock) in the ISM $c_{\rm s}$, ${\cal M}_{\rm s}  = {\rm v}_{\rm s} /c_{\rm s}$. From the standard theory of the gas with an adiabatic index $\gamma = 5/3$,  the compression ratio at the shock is expressed as (Landau \& Lifshic \cite{ll86})
\begin{equation}
x_{\rm s}  = \frac{{4{\cal M}_{\rm s}^2 }}{{{\cal M}_{\rm s}^2  + 3}};
\label{eq3}
\end {equation}
and for the dependence of the spectral index $q$ on the Mach number, we have 
\begin{equation}
q  = \frac{{4{\cal M}_s^2 }}{{{\cal M}_s^2 - 1}}.
\label{eq4}
\end{equation}

It is important to note that it is much easier to apply the mechanism of DSA to the evolved SNRs than to the young ones.
This is 
 first because it is expected that various nonlinear and geometrical effects do not play an essential role in the evolved SNRs, so they can be neglected and the test particle approximation is acceptable. Second, the structure of the shock wave of the evolved SNRs in homogeneous ISM can be described in a simple way, unlike the young ones. In the young SNRs the shape of the circumstellar matter distribution often plays a crucial role, and  they often  have a double- or even multiple- shell structure. Third, it is expected that in the evolved SNRs the temperature equipartition between electronic and proton components takes place. 

One very important problem remaining still unsolved in the theory of DSA is the  injection of particles into the action of the mechanism. Although in the case of protons, there is some progress in this direction (e.g., Drury \& Malkov \cite{maldr}), the theory of injection practically is not advanced for electrons. One basic theoretical difficulty in applying DSA to electrons is that it is still unclear how thermal electrons gain momentum above the thermal ions, so they can be injected into a population that can respond to DSA. In our case, the determination of the coefficient $A$ in the spectrum (\ref{eq1}) is directly connected to this injection problem. As initially done by Bell (\cite{b78b}) in applications of DSA to real objects, the injection problem is usually avoided by assuming that some fraction of the thermal particles are ``injected'' as non-thermal particles at some  ``injection momentum'', $p_{\rm inj}$, which separates the thermal particle population from the non-thermal. Here we introduce these two parameters as 

\begin {equation}
\eta  = N_{\rm inj} /n_{\rm e} 
\label{eq5}
\end{equation}	
\begin {equation}
\psi  = p_{\rm inj} /p_{\rm th}
\label{eq6}
\end{equation}
where  $N_{\rm inj}$ is the concentration of electrons contained in the spectrum (\ref{eq1}), $p_{\rm inj}$ the injection momentum, and $p_{\rm th}$ the thermal momentum of heated by the shock plasma with number density $n_{\rm e}$.

In present paper we assume that suprathermal electrons from the tail of the downstream Maxwellian distribution function are  subjected to acceleration, which allows us to determine the coefficient in the spectrum of the accelerated electrons through the equality

		\[ \int\limits_{p_{{\rm inj}} }^{p_{{\rm max}} } {4\pi p^2 f_{\rm a} (p)dp}  = \int\limits_{p_{{\rm inj}} }^\infty  {4\pi p^2 f_{\rm m} (p)dp}\,\,\,, \]
where		\[f_{\rm m} (p) = \frac{{n_{\rm e} }}{{\pi ^{3/2} p_{\rm th}^3 }}\exp \left( { - \frac{{p^2 }}{{p_{\rm th}^2 }}} \right)\]  is the Maxwellian distribution function of electrons heated and compressed by the shock, and $p_{\rm th}$  is the thermal momentum of electrons $p_{\rm th}  = (2m_{\rm e} T_{\rm s})^{1/2}$. We also assume that there is an equilibrium between electron and proton temperatures, i.e. $T_{\rm{e}}  \cong T_{\rm{s}}  = (3/16)\,\bar \mu \,m_{\rm p} \rm{v}_{\rm{s}}^{\rm{2}}$  ($\bar\mu$ is the mean mass per particle in units of the proton mass), which is expected to be the case in evolved SNRs. 
From the above equality, we have for $A$ an expression

\begin{equation}
A = \frac{{\eta n_{{\rm e0}} }}{{4\pi (1 - \varepsilon ^{q - 3} )}}qp_{{\rm inj}}^{q - 3} 
\label{eq7}
\end {equation}
where ${\rm  }\varepsilon  = p_{{\rm inj}} /p_{{\rm max}}$,
\begin{equation}
\eta  \equiv \frac{{N_{{\rm inj}} }}{{n_{\rm e} }} = \frac{4}{{\sqrt \pi  }}\int\limits_\psi ^\infty  {x^2 e^{ - x^2 } dx} 
\label{eq8}
\end {equation}
with $\eta  \approx 4.60 \times 10^{ - 2}, \,\,5.85 \times 10^{ - 3}, \,\,4.40 \times 10^{ - 4}$ for  $\psi \equiv p_{{\rm inj}}/p_{{\rm th}} = 2,\,\,2.5,\,\,3$, respectively.
 In deriving expression (\ref{eq7}) we have used $n_{\rm e}  = x_{\rm s}\, n_{{\rm e0}}$ from the shock jump conditions and $q/(q - 3)  = x_{\rm s}$, which follows from (\ref{eq2}).

Taking (\ref{eq7}) into account, we can rewrite the spectrum of accelerated electrons  as
\begin{eqnarray}
f_{\rm a} = \frac{{\eta \,n_{{\rm e0}} }}{{4\pi \,p_{{\rm inj}}^3 (1 - \varepsilon ^{q - 3} )}}\,q\left( {\frac{p}{{p_{{\rm inj}} }}} \right)^{ - q}  \nonumber\\\times 
\left[ {H\left( {p - p_{{\rm inj}} } \right) - H(p - p_{{\rm max}} )} \right]
\label{eq9}
\end {eqnarray}
It is interesting to note that this distribution function has the same form as the one derived by using the $\delta $ - function as the source of injected particles. 

This elementary ``solution'' of the problem of injection has some theoretical basis. Indeed, theoretical works devoted to the problem of electron injection and following acceleration in various approximations show that the overall spectra of electrons has modified Maxwellian distributions with power-law tails at high energies (e.g., Bykov \& Uvarov \cite{byk99}). The hypothesis used above about the injection of particles directly from the Maxwellian distribution is not entirely physically   correct.   Only under very specific conditions can superthermal electrons be accelerated by DSA, even though in this case our consideration is very simplified. The problem of particle acceleration from the Maxwellian distribution was considered in detail by Gurevich (\cite{gurevich})  (see Bulanov \& Dogiel \cite{buldog} for the case of shock acceleration and Dogiel  \cite{dog}), who showed that deviation from the equilibrium Maxwellian distribution begins at the momenta much lower than $p_{{\rm inj}}$, and there is broad transition region between the equilibrium Maxwellian distribution and power-law spectrum at high energies. As shown by Asvarov et al. (\cite{asv90}), this part of the electron spectrum may manifest itself through the bremsstrahlung X-ray emission at photon energies $\ga 10 ~{\rm keV} $. Indeed, such emission has been detected from several SNRs, although most authors interpret it as synchrotron emission from  TeV-energy electrons; however, such an interpretation cannot be considered as  proved,  especially in the case of   Cas A  SNR (e.g., Vink \cite{vink}). Physically it is most probable that the low energy suprathermal electrons are accelerated first by the electric fields (or stochastically) and then by DSA to the higher energies. Although it is difficult to offer any concrete mechanism accelerating electrons from the thermal distribution up to the energies sufficient for DSA, but our simple assumption can be considered as the first approximation, and it is more or less good if  $N_{{\rm inj}} \ll n_{{\rm 0e}}$.  And finally, if we accept this naive representation about injection then the two Bell parameters turn out to be related to each other through relation (\ref{eq8}).

The distribution function (\ref{eq9}) is derived for the plane and steady shock, neither of which is fulfilled in the case of SNRs. %' blast wave shocks. 
To take the effects of nonstationarity and sphericity into consideration and to follow the evolution of the spectrum of accelerated particles in a simple way, it is useful to employ the method known as ``onion-shell'' (Moraal \& Axford \cite{morax}; Bogdan \& V\"olk \cite{bovo83}). For our purposes, the method of Moraal \& Axford (\cite{morax}) is more appropriate, so we use it with some modifications. 

First of all, we assume that at any moment of interest the spectrum of electrons (\ref{eq9}) extends up to the energies (say, $ ~30\,\,{\rm GeV}$)  sufficient for emitting in the range of radio frequencies $(0.1\div 10)\,\,{\rm {GHz}}$ for the plausible values of the magnetic fields between ${\rm 3 }~\mu {\rm G}$  and ${\rm 10}^{{\rm  - 4}} \,{\rm G}$, in other words, in formula (\ref{eq9}) we consider that $p_{{\rm max}}  \to \infty$ ($\varepsilon=0$). 

 Taking these assumptions into account and expressing the injection momentum as a function of shock Mach number $p_{{\rm inj}}  = p_{{\rm 0i}} ({\cal M}_{\rm s} /{\cal M}_{\rm 0})$, where $p_{\rm 0i}$ is the injection momentum at the initial value of the Mach number, ${\cal M}_{\rm 0}$,  when the shock radius was $R_{\rm 0}$, the spectrum (\ref{eq9}) takes the form
\begin {equation}
f_{\rm a}  = \frac{{\eta \,\,n_{{\rm e0}} }}{{4\pi \,p_{{\rm 0i}}^3 }}\,q\,\left( {\frac{p}{{p_{{\rm 0i}} }}} \right)^{ - q} \left( {\frac{{{\cal M}_{\rm s} }}{{{\cal M}_{\rm 0} }}} \right)^{q - 3}\, H\left( {p - p_{{\rm 0i}}\, {\frac{{\cal M}_{\rm s}}{{\cal M}_{\rm 0} }}} \right)
\label{eq10}
\end {equation}	

Let us introduce the radially dependent distribution function, $f(p,R,R_{\rm s})$,  which will allow us to find the concentration of accelerated particles at any radial position inside the SNR with a given radius $R_{\rm s}$. To do this, one must first adopt a concrete model describing the structure of the SNR. As was noted before for the model of the SNR, we use an analytical approximation of  Cox \& Andersen (\cite{coxa82}) (details of this approximation are given in Appendix). Second, we assume that the accelerated electrons in the blast wave shell are frozen-in to the matter and participate in the bulk motion of the gas without any dynamical effect; i.e. there is no diffusion of electrons. This is not a very strong constraint for the value of the diffusion coefficient, if we take into account that in the gyroradius limit for $\sim 1$~GeV electrons in the $B\sim 10^{ - 5}$~G magnetic fields $\kappa _{{\rm min}} {\rm  } \cong {c^{2}}p/3eB \sim 3\times 10^{21} \,\,{\rm cm}^{\rm 2} {\rm /s}$,  and there is observational evidence that the actual value of the diffusion coefficient in the region of the shock front is only a few times larger than $\kappa_{\min }$ (Achterberg et al. \cite{ABR94}). Finally, we consider adiabatic cooling of the accelerated electrons in the course of  expansion of the shell. 
The number of particles, 
$dN$, in the momentum range $(p;p + dp)$ contained inside the SNR with $R_{\rm s}$ is
\begin {equation}
dN = 4\pi p^2 dp \, \int\limits_{R_{{\rm in}} }^{R_{\rm s} } {4\pi R^2 dR \, f(p,R,R_{\rm s} )}, 
\label{eq11}
\end {equation}
where $R_{\rm in}$ is the new location of a mass element that was first shocked at $R_{\rm 0}$. We assume that there are no particles inside the radius $R_{\rm in}$, which means that the action of the acceleration mechanism first starts when the shock radius is $R_{\rm 0}$. 

The same number of particles, $dN$, equals the sum of particles that were initially accelerated into the momentum range $(p_{\rm i},p_{\rm i} + dp_{\rm i})$  and were contained in the compressed volume $dV_{\rm i}  = 4\pi R_{\rm i}^2 dR_{\rm i} /x_{{\rm si}}$  ($x_{\rm si}$  is the compression ratio at the moment when shock radius was $R_{\rm i}$) and subsequently cooled down to $(p;p + dp)$. To take  the adiabatic cooling into account, we introduce a cooling function as in (Moraal \& Axford \cite{morax})
\[C = p/p_{\rm i}  = \left( {\frac{1}{{x_{{\rm si}} }}\frac{{dV_{\rm i} }}{{dV}}} \right)^{1/3}, \]
which describes loss of particle momentum from $p_{\rm i}$  to $p$  due to expansion of the mass element from $1/\rho _{\rm i}$  to $1/\rho$. Using  (\ref{A3}) and (\ref{A4}), the cooling function takes the form:
\begin{equation}
C = \left( {x(r,R_{\rm s} ) /x_{{\rm si}} } \right)^{1/3}. 
\label{eq12}
\end{equation}
Since the isotropic distribution function remains unchanged by this type of cooling, now we can find $dN$ by integrating over all volumes $dV_{\rm i}$
\begin{equation}
dN = 4\pi p^2 dp\int\limits_{R_{\rm 0} }^{R_{\rm s} } {4\pi R_{\rm i}^2 dR_{\rm i} \cdot \frac{{f_{\rm a} (p/C,R_{\rm i} )}}{{x_{{\rm si}} C^3 }}}
\label{eq13} 
\end{equation}
where  $C$ depends on $R_{\rm i}$ and $R_{\rm s}$. Equating (\ref{eq11}) and (\ref{eq13}), we have 
\[\int\limits_{r_{i{\rm n}} }^1 {r^2 dr \cdot f(p,r,R_{\rm s} )}  = \int\limits_{R_{\rm 0} /R_{\rm s} }^1 {r_{\rm i}^2 dr_{\rm i} }  \cdot \frac{{f_{\rm a} (p/C,r_{\rm i} )}}{{x_{{\rm si}} C^3 }}\]
where we introduced notations for the normalized radii as $r=R/R_{\rm s}$  and $r_{\rm i}=R_{\rm i}/R_{\rm s}$.  Using the expressions (\ref{A3}) and (\ref{A4}), after transformation 
$r_{\rm i}  = r_{\rm i} (r)$,  $dr_{\rm i}  = \frac{{dr_{\rm i} }}{{dr}}dr = x(r) \, \frac{{r^2 }}{{r_{\rm i}^2 }}dr$  in the right hand side integral and taking (\ref{eq10}) into account, we obtain
\begin{eqnarray}
 f(p,r,R_{\rm s} ) = \frac{{\eta \,n_{{\rm 0e}} }}{{4\,\pi \,p_{{\rm 0i}}^3 }}\, q_{\rm i} \,C^{q_{\rm i}} \left( {\frac{p}{{p_{{\rm 0i}} }}} \right)^{ - q_{\rm i}} \nonumber\\
 \times 
\left( {\frac{{{\cal M}_{\rm i} }}{{{\cal M}_{\rm 0} }}} \right)^{q_i  - 3} H (r - r_{{\rm in}} )\,H \left({p - Cp_{\rm 0i} \frac{{{\cal M}_{\rm i} }}{{{\cal M}_0 }}} \right).
\label{eq14}
\end {eqnarray}
In this spectrum, $q_{\rm i}= q_{\rm i}(r,{\cal M}_{\rm i})$ is the spectral index of particles that were accelerated earlier when the shock front was at radius $R_{\rm i}$ with strength  ${\cal M}_{\rm i}$ and that are now at $R=r\,\,R_{\rm s}$  inside the SNR with radius $R_{\rm s}$. For any radial position  $r$ at the present shock strength ${\cal M}_{\rm s}$,  the value of  ${\cal M}_{\rm i}$  can be found by solving the equation (see Appendix 1)
\begin{equation}
\frac{{\left( {{\cal M}_{\rm i}^2  - 11/15} \right)^{2/3} }}{{({\cal M}_{\rm i}^2  - 1)}} = r_{\rm i} (r,{\cal M}_{\rm s} )\,\, \frac{{\left( {{\cal M}_{\rm s}^2  - 11/15} \right)^{2/3} }}{{({\cal M}_{\rm s}^2  - 1)}}.
\label{eq15}
\end {equation}
The value of inner radius $r_{\rm in}$  in (\ref{eq14}) for any $R_{\rm s}$  (and ${\cal M}_{\rm s}$, respectively) can be found as the root of Eq. (\ref{eq15}) but with substitutions of ${\cal M}_{\rm 0}$ for ${\cal M}_{\rm i}$ and $r_{\rm in}$  for $r$.

Now if we write the distribution function in the form 
\begin {equation}
dN = K_{\rm e} E_{\rm k}^{ - (q - 2)} dE_{\rm k} \,\,\,{(\rm electron/cm}^{\rm3}), 
\label{eq16}
\end {equation}
than taking into account an equality $K_{\rm e} E_{\rm k}^{ - (q - 2)} dE_{\rm k}  = 4\pi p^2 dp \, f(p,r,R_{\rm s} )$  and that $E \cong pc$  for the range of energies of interest for the coefficient in (\ref{eq16}), we obtain  an expression
\begin{eqnarray}
K_{\rm e}  = \eta n_{{\rm e0}}\,(p_{\rm 0i} c)^{q_{\rm i}  - 3}\,q_{\rm i} C^{q_{\rm i}} \, \left( {\frac{{{\cal M}_{\rm i}}}{{{\cal M}_0 }}} \right)^{q_{\rm i}  - 3} \nonumber\\
 \times
  \,\,\,H (r - r_{\rm in})\,H \left({p - Cp_{\rm 0i} {{\cal M}_{\rm i} }/{{\cal M}_0 }} \right).
\label{eq17}
\end {eqnarray}

\subsection{The coefficient of synchrotron emission.}

Isotropically distributed electrons with the distribution function (\ref{eq16}) in the fully chaotic magnetic field $B$ radiate synchrotron radio emission with the coefficient (Ginzburg \cite{g81})
\begin{eqnarray}
\varepsilon _\nu   = 1.35 \times 10^{ - 22} \,a(\alpha )\,K_{\rm e} B^{\alpha  + 1} \left( {\frac{{6.26 \times 10^{18} }}{\nu }} \right)^\alpha  
 \nonumber\\
{(\rm erg\,\,cm}^{\rm -3} \,{\rm sr^{\rm -1}}\,{\rm Hz^{\rm -1})}
\label{eq18}
\end{eqnarray}
where  $\nu$ is the frequency; $\alpha$  is the radio spectral index, which is connected with the value of the spectral index of the particle distribution, $q$, as $\alpha=(q-3)/2$, $a(\alpha)$ is the coefficient fully depending on the value of $\alpha$  (Ginzburg \cite{g81}; p.93). Assuming the magnetic field is dependent on the density as $B \propto \rho ^{k}$,  we represent its radial distribution inside the remnant as

\begin{equation}
B = B_{\rm 0}  \cdot \left[ {x(r,{\cal M}_{\rm s} )} \right]^{k} 
\label{eq19}
\end{equation}
where $B_{\rm 0}$  is the ISM magnetic field strength, and $x(r,{\cal M}_{\rm s})$ the radial density distribution (\ref{A3}).

Taking the definitions (\ref{eq6})--(\ref{eq8}) into account, the badly determined parameter of $p_{\rm 0i}$ can be expressed  as $(p_{{\rm 0i}} c)^2  = (\psi \,p_{{\rm T}_{\rm 0} } c)^2  = \psi ^2  \cdot 2m_{\rm e} c^2 \,T_{{\rm 0s}}$, with $T_{{\rm 0s}}  = 3.13 \times 10^{ - 7} \bar \mu \,{\rm v}_{{\rm 09}}^2 \,\,({\rm erg)}$, where ${\rm v}_{{\rm 09}} = {\rm v}_{{\rm 0s}} /(10^9 \,{\rm cm/s})$ is the initial shock velocity when the shock radius was $R_{\rm 0}$. After substituting (\ref{eq17}) into (\ref{eq18}) and taking the last considerations concerning $B$  and $p_{\rm 0}$ into account, we obtain the expression for the emission coefficient in the form
\begin{equation}
\varepsilon _\nu = \varepsilon _{{\rm \nu 0}} \cdot \varepsilon_\nu (r)
\label{eq20}
\end{equation}
where
\begin{eqnarray}
\varepsilon _{\nu 0} = 9.97 \times 10^{ - 35} \,\eta _{{\rm  - 3}} \,n_{{\rm e0}}\, B_{{\rm  - 5}} \,\,2^{3k}\, c_1 ^{0.5} 
  \nonumber\\ 
 \,\,\,\,{\rm (erg}\,\,{\rm cm}^{{\rm  - 3}} {\rm sr}^{{\rm  - 1}} {\rm  Hz}^{{\rm  - 1}} {\rm )},
\label{eq21}
\end{eqnarray}
and
\begin{eqnarray}
\varepsilon_\nu (r) = \frac{{a(\alpha _i )}}{{a(0.5)}}\,\frac{q_{\rm i}}{4} \,C^{q_{\rm i}}\frac{{[x(r)]^{k (\alpha _{\rm i}  + 1)} }}{{2^{3k} }}
\, \nonumber\\ \times 
\left( {3.21 \cdot 10^{ - 8}\, c_1 }\right)^{\alpha _{\rm i}  - 0.5}\,
\left( {\frac{{{\cal M}_{\rm i}}}{{{\cal M}_{\rm 0} }}} \right)^{2\alpha _{\rm i}}\,H(r - r_{{\rm in}} ),
\label{eq22}
\end{eqnarray}
where $c_{\rm 1}  =\bar\mu \,\psi ^2 \,{\rm v}_{{\rm 09}}^2 \,B_{{\rm  - 5}} /\nu _9 $ ,
 $\eta_{\rm -3}$  is $\eta$ in units of $10^{-3}$, $B_{\rm -5}$ is $B_{\rm 0}$ in units of  $10^{-5}\,\,{\rm G}$ and $\nu_{\rm 9}$ is the radio frequency in units of $10^{9}\,\,{\rm Hz}$. The radially dependent part of the emission coefficient (\ref{eq22}) is constructed in such a way that at $r=1$ and ${\cal M}_{\rm s}={\cal M}_{\rm 0}$ it becomes unity. It is important to note that the poorly known parameter $\psi $ enters the formula for emissivity through $c_1$ as a combination with other parameters and, as shown below, the dependence of the evolutionary part of the emission coefficient on $c_1$ is very weak.

Using (\ref{eq20})-(\ref{eq22}), the formula for the radio flux density can be written  as
\begin{equation}
S_\nu   = S_{0\nu }  \cdot l_\nu  (R_{\rm s} )
\label{eq23}
\end{equation}
where
\begin{equation}
S_{0\nu }  = 387\,\,\eta _{{\rm  - 3}} \,\,n_{{\rm e0}} \,\,2^{3k} \,B_{{\rm  - 5}} \,\,c_{\rm 1}^{0.5}  \cdot R_{{\rm 0pc}}^3 \,d_{{\rm kpc}}^{ - 2} \,\,{\rm (Jy)}
\label{eq24}
\end{equation}
and
\begin{equation}
l_\nu  ({R}_{\rm s} ) = \left( {\frac{{R_{\rm s} }}{{R_{\rm 0} }}} \right)^3  \cdot \int\limits_{r_{{\rm in}} }^1 {\varepsilon_\nu (r) \cdot r^2 dr} 
\label{eq25}
\end{equation}
and the surface brightness
\begin{equation}
\Sigma _{\rm \nu }  = \Sigma _{{\rm \nu 0}}  \cdot \sigma_{\rm \nu } (R_{\rm s} )
\label{eq26}
\end{equation}
where
\begin{eqnarray}
\Sigma _{0\nu }  = 1.23 \times 10^{ - 18} \,\eta _{{\rm  - 3}} \,n_{{\rm e0}} \,\,2^{3k} \,B_{{\rm  - 5}} \,c_{\rm 1}^{0.5}R_{0pc}^{}  \nonumber\\
 \left( {{\rm W}\,\,{\rm m}^{ - {\rm 2}} {\rm sr}^{ - {\rm 1}} {\rm Hz}^{ - {\rm 1}} } \right)
\label{eq27}
\end{eqnarray}
and
\begin{equation}
\sigma _\nu  ({R}_{\rm s} ) = \left( {\frac{{R_{\rm s} }}{{R_{\rm 0} }}} \right) \cdot \int\limits_{r_{{\rm in}} }^1 {\varepsilon_{\nu} (r) \cdot r^2 dr}. 
\label{eq28}
\end{equation}
In (\ref{eq24}) and (\ref{eq27}), $R_{\rm 0pc}$ is the initial radius of SNR in $\rm pc$ and $d_{\rm kpc}$  is the distance to SNR in kpc. The ratio $R_{\rm s} /R_{\rm 0} $ in (\ref{eq25}) and (\ref{eq28}) can be expressed through the Mach number by using the approximated formula (\ref{A2}).

\subsection{Common properties of the model}
In this section we study general properties of the model, namely, the evolution of the main radio characteristics (emissivity, luminosity, surface brightness, spectral index) of the SNR with Mach number and their dependence on the input parameters of the model. The Mach number is used as the parameter describing the evolution of the SNR although it can be easily transformed into other measures of evolution such as radius and/or time by using Eqs (\ref{A2}) and (\ref{A6}). There are three input model parameters: initial Mach number ${\cal M}_{\rm 0}$, the power $k$   in the relationship $B \propto \rho^k$,  and the combination of parameters $c_{\rm 1}  =\bar \mu\, \psi ^2 {\rm v}_{09}^2  \cdot B_{{\rm - 5}} /\nu _9$.  %
Both, ${\cal M}_{\rm 0}$  and $c_{\rm 1}$ are not fully independent, but in this subsection we adopt them as independent parameters. The value of $c_{\rm 1}$ is restricted within the range between $\approx 1$ and $\approx 200$ when the variables $\rm v_{09}$ and $B_{{\rm - 5}}$ are changing in the range $0.5 - 5$ and $0.3 - 1$, respectively. It is important to note that, as shown below, the main radio properties of the SNR depend very weakly  on the value of $c_{\rm 1}.$

\begin{figure}
\centering
\includegraphics[width=7.5cm]{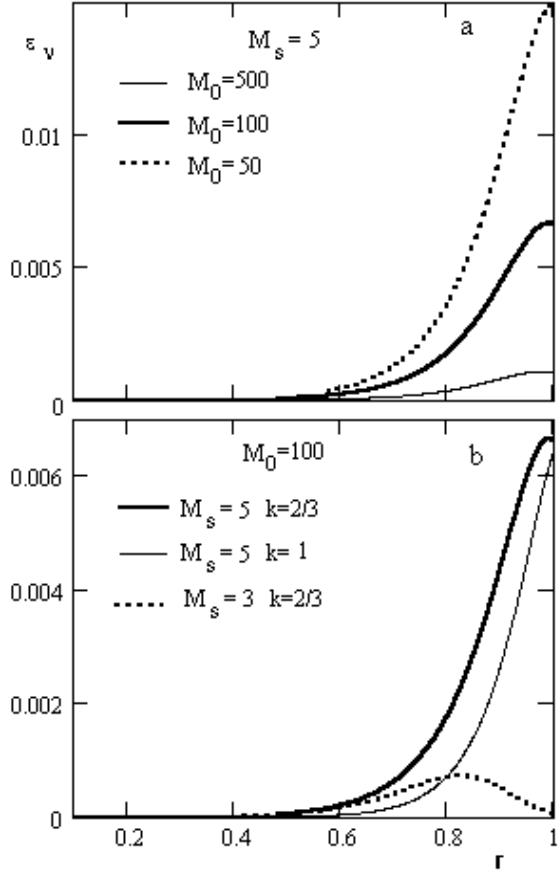}
\caption{The evolution of the radial profiles of an undimensional coefficient of  emission $\varepsilon_\nu (r)$ (Eq.\ref{eq22}) in its dependence on the input parameters.}
\label{fig1}
\end{figure}

In Fig.~\ref{fig1}a the normalized radial profiles of the volume emissivity coefficient, $\varepsilon_\nu (r)$, are shown for different sets of input parameters. As seen in this figure, emission at high Mach numbers is highly concentrated directly behind the shock front; but with decreasing the shock strength, the profiles broaden and, ultimately, at some critical Mach number, ${\cal M}_{\rm cr}$, the peak of the profile separates from the shock front. This occurs around ${\cal M}_{\rm cr}  \la 5$ when the decrease in the efficiency of acceleration due to steeper spectrum of the newly accelerated electrons becomes more effective than the adiabatic cooling of early accelerated electrons. As the magnetic field is frozen well in the ionized plasma, its strength reaches its maximum directly behind the shock front, but the peak in the distribution of radio emitting electrons is displaced from the regions where the magnetic field strength is at its maximum. Thus, the effectiveness of transforming of the energy of particles and magnetic fields into the synchrotron emission decreases (Asvarov \& Guseinov \cite{asvgus}).
%The role of this effect strongly depends on the form of the dependence between the magnetic field strength and the density of an ionized gas (Fig.\ref{fig1}b).

\begin{figure}
\centering
\includegraphics[width=7.5cm] {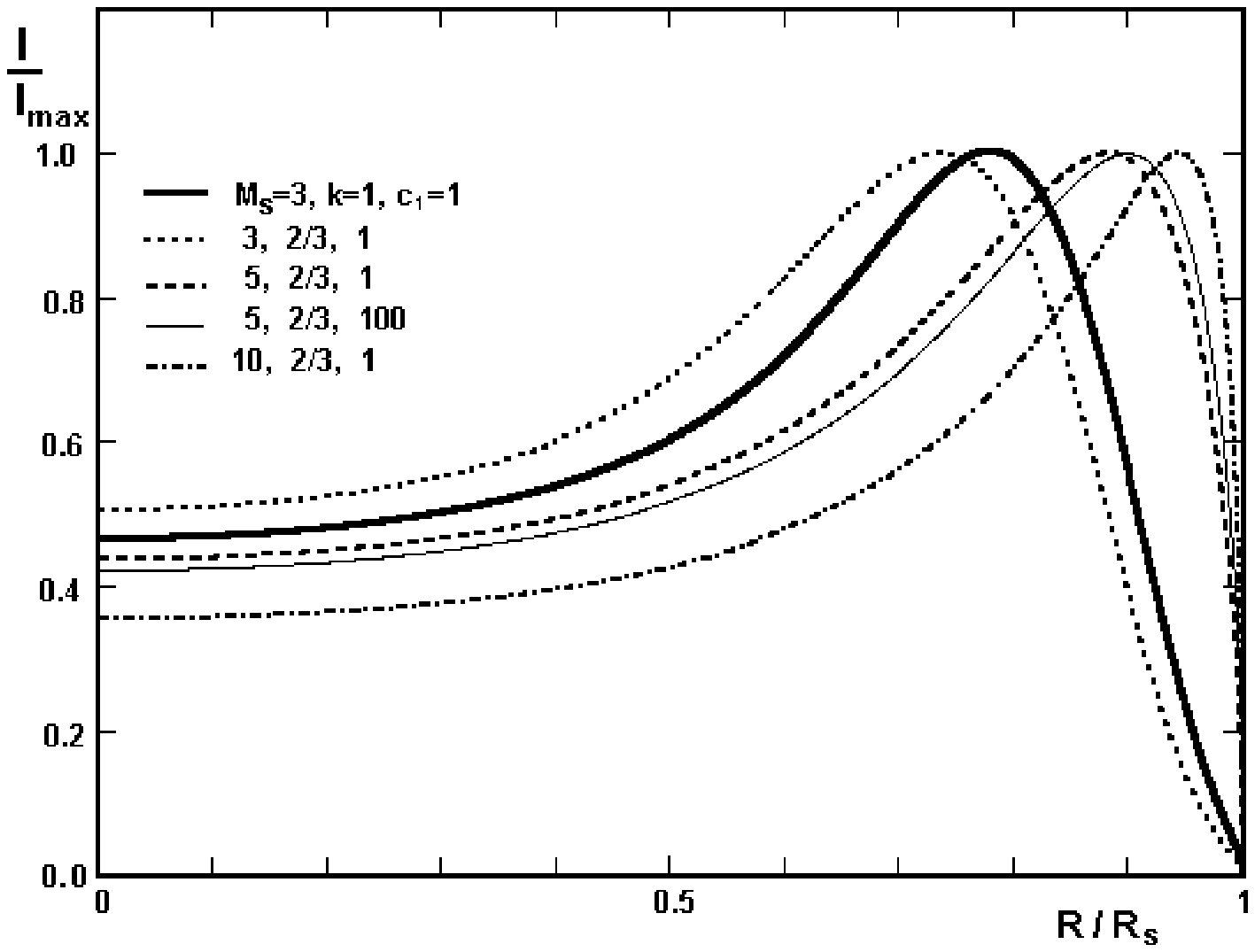}
\caption{Normalized projected radial profiles of synchrotron emission for ${\cal M}_{\rm 0}  = 100$  and different sets of parameters $({\cal M}_{\rm s} ,k ,c_{\rm 1} )$.}
\label{fig2}
\end{figure}

 The effect of broadening the profile is more prominent for high values of ${\cal M}_{\rm 0}$  and small $c_{\rm 1}$  at fixed other parameters. The radial distributions of emissivity presented in Fig.\,\ref{fig1} lead to the radial distributions of the radio intensity on the face of SNR shown in Fig.~\ref{fig2} for several set of input parameters calculated by integration along the line of sight $I(z) = \int {\varepsilon_\nu (r) \cdot dl}$, where $z$ is the radial position on the radio image of SNR. 

The main features that can be seen in this figure are: \\ 1) In evolved SNRs the radius of the radio remnant defined as the radius of the circle of the maximum intensity  is smaller than its real radius, though for young remnants the peak of the intensity occurs directly behind the shock front. This implies that for old  adiabatic  SNRs with ${\cal M}_{\rm s}  < 10$ the radio diameter systematically smaller then its real diameter. The difference between these diameters increases with decreasing shock Mach number. \\ 2) As can be seen in Fig.~\ref{fig2}, broadening of the radio shell in the course of the evolution of the SNR is the next property of the model. \\ 3) The center-to-limb ratio of the projected profile of radio emission is the measurable characteristic of an SNR. Our model predicts for this ratio a value of $ \la 0.4$ at high Mach numbers  and  increases up to a value of $ \sim 0.5$ with decreasing $ {\cal M}_{\rm s}$. At given ${\cal M}_{\rm s}$ and ${\cal M}_{\rm 0}$ the dependence of the shape of $I(z)-$ profiles on the other input parameters is weak.

\begin{figure}
\centering
%\resizebox{\hsize}{!}{\includegraphics{Fig4Fin.eps}}
\includegraphics[width=7.5cm] {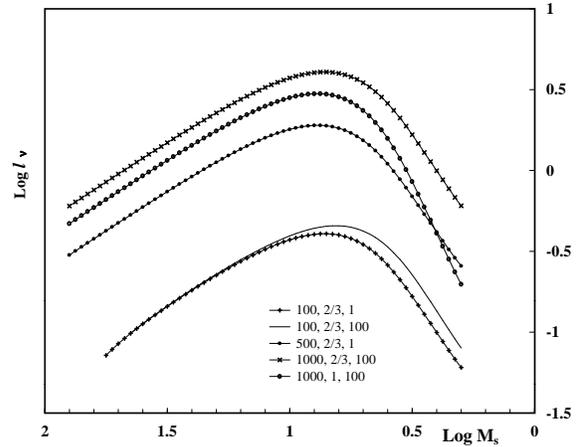}%
\caption{The dependence of undimensional luminosity (Eq. ~\ref{eq25}) on ${\cal M}_{\rm s}$. The tracks are labeled with input parameters $({\cal M}_{\rm 0}, \,k, \,c_{\rm 1} )$ }
\label{fig3}
\end{figure}

  Figure ~\ref{fig3} shows the evolution of the undimensional luminosity $l_\nu $   with ${\cal M}_{\rm s}$ for several sets of input parameters $({\cal M}_{\rm 0}; \,k; \,c_{\rm 1} )$. This quantity at first increases as   
$l_\nu \propto {\cal M}_{\rm s}^{-1}$ with decreasing Mach number; then after reaching its maximum at ${\cal M}_{\rm max}$, it decreases as  $l_\nu \propto {\cal M}_{\rm s}^{2.2}$ when $k = 2/3$ and as $\sim {\cal M}_{\rm s}^{3}$ for $k = 1$. The corresponding dependencies of $l_{\nu}$ on the shock radius ${R_{\rm s}}$ also can be approximated by two power-law relationships: $l_\nu \propto R_{\rm s}^{1.4}$ and $l_\nu \propto R_{\rm s}^{-2.6} (k=2/3)$ or $ \sim R_{\rm s}^{-3.2} (k=1)$. The values of ${\cal M}_{\rm max}$ and the corresponding peak values of the undimensional luminosity $l_{\nu \rm max}$ are given in Table ~\ref{tabl1}.  The value of $l_{\nu \rm max}$ linearly increases with ${\cal M}_{\rm 0}$ but the dependence on the parameter $c_{\rm 1}$ is weak.   

\begin{figure}
%\centering
%\resizebox{\hsize}{!}{\includegraphics{Fig4Fin.eps}}
\includegraphics[width=7.5cm] {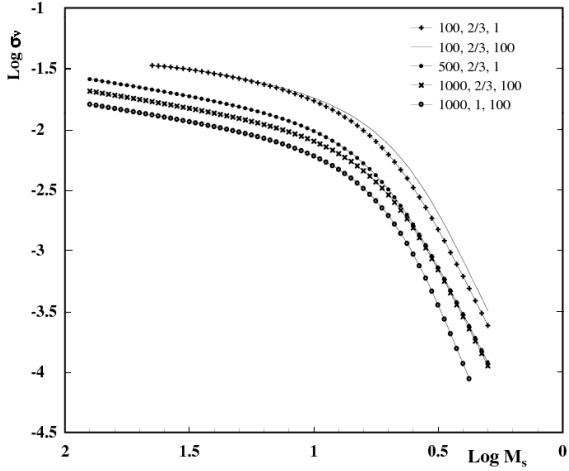} %{Fig4N.eps}
\caption{The dependence of undimensional surface brightness (Eq. ~\ref{eq28}) on ${\cal M}_{\rm s}$. Curves are labeled with input parameters $({\cal M}_{\rm 0}, \,k, \,c_{\rm 1} )$ }
\label{fig4}
\end{figure}

Figure~\ref{fig4} shows the evolution of undimensional surface brightness, $\sigma_\nu$, (see Eq. ~\ref{eq28}) with ${\cal M}_{\rm s}$ for several set of input parameters. As follows from the analysis of tracks, the relationship $\sigma_{\nu} - {\cal M}_{\rm s}$   has a universal character within the framework of the adopted model. The dependence $\sigma_\nu = \sigma_\nu ({\cal M}_{\rm s})$ can also be approximated by the two power-law dependencies: $\sigma_\nu  \propto {\cal M}_{\rm s}^{0.3 \div 0.4}$ at ${\cal M}_{\rm s}  \geq {10}$ and $\sigma_\nu  \propto {\cal M}_{\rm s}^{4 \div 5}$  at ${\cal M}_{\rm s}  \leq {5 - 6}$. The transition to  declining branch is smooth and begins at ${\cal M}_{\rm s}\leq 10$. The values of $\sigma_\nu$ at ${\cal M}_{\rm s}=10$  and ${\cal M}_{\rm s}={\cal M}_{\rm max}$ are given in Table~\ref{tabl1}.  At low values of ${\cal M}_{\rm s}$,  the  slope of the curve $\sigma_\nu = \sigma_\nu ({\cal M}_{\rm s})$  is steeper when the degree of relationship between the magnetic field and the density is stronger (curve with $k=1$ in Fig.~\ref{fig4}).  At high ${\cal M}_{\rm s}$, the value of $\sigma_\nu$ depends weakly on ${\cal M}_{\rm 0}$ and $k$ and barely depends on parameter $c_1$. The ratio of the extreme values of $\sigma_\nu$  calculated for the set of input parameters within their maximum plausible boundaries is a factor of 2.5 at ${\cal M}_{\rm s} =10$  and does not exceed a factor of 3.5 for ${\cal M}_{\rm s } > 4$, whereas this ratio for $l_\nu $ is much higher (see, Table ~\ref{tabl1}). 

\begin{table}
\caption{The parameters of the model. $\sigma_{\rm max}$ is $\sigma_\nu$ at ${\cal M}_{\rm max}$, $\sigma_{\rm 10}$ is $\sigma_\nu$ at ${\cal M}_{\rm s}=10$ } 

\label{tabl1}
%\medskip
\centering\begin{minipage}{8.5cm}
\begin{tabular}{rcccccc} \hline
\hline
${\cal M}_{\rm 0}$ & $k$  & c$_{\rm 1}$ & ${\cal M}_{\rm max}$ &$l_{\nu \rm max}$&$\sigma_{\rm max}$& $\sigma_{\rm 10}$   \\ 
\hline
 &   &  &  & & $\times 10^{ - 3}$ &  $\times 10^{ - 3}$  \\ 
\hline
   1000 & 2/3 & 1  &7.9  &3.710 &5.734&7.532\\ 
   1000 & 2/3 & 100&7.2  &4.066 &5.542&7.958\\ 
   500  & 2/3 & 1  &7.7  &1.902 &7.161&9.647\\ 
   500  & 2/3 & 100&6.9  &2.098 &6.795&10.19\\ 
   100  & 2/3 & 1  &7.2  &0.406 &11.93&17.12\\ 
   100  & 2/3 & 100&6.5  &0.455 &11.60&18.10\\ 
   100  & 1   & 1  &7.8  &0.299 &9.780&12.94\\ 
   100  & 1   & 100&6.9  &0.334 &9.244&13.75\\ 
\hline
\end{tabular}
\end{minipage}
\end{table}

Although this is a very important prediction of the model, the Mach number is not an observable parameter. Remnants with the same values of ${\cal M}_{\rm s}$ may have completely different diameters. 
Converting the dependence of the surface brightness on the Mach number into the surface brightness - diameter dependencies, we find that at ${\cal M}_{\rm s} \ga 10$ surface brightness changes with SNR radius as $R_{\rm s}^{ - (0.3 \div 0.5)}$, followed by a steeper decrease with radius as $R_{\rm s}^{ - (4.5\div5)}$ at ${\cal M}_{\rm s}  \la 8$. This transition  between two power laws is  smooth enough, because it begins at ${\cal M}_{\rm s} \ga 10$ and finishes at ${\cal M}_{\rm s} \sim 5$.  Moreover, if we take into account the above-mentioned effect of lagging of the radio shell behind the main shock front, the slope will be even  steeper.  The prediction of the model about the "surface brightness--diameter" relationship agrees with the conclusion of Berkhuijsen (\cite{ber86}) that SNRs evolve at a constant surface brightness followed by an abrupt drop, although our conclusion is true if the evolution of SNR takes place without the onset of the radiative phase. The case with possible radiative cooling is considered below.
Our result disagrees with the conclusion of Berezhko \& V\"olk (\cite {berezhko}) that in the Sedov phase $\Sigma_{\rm R} \propto R_{s}^{-17/4}$, which  is evidently the result of the assumption made by  these authors about generation of the magnetic field at the shock front.  

\begin{figure}
\centering
%\resizebox{\hsize}{!}{\includegraphics{Fig5.eps}}
\includegraphics[width=7.5cm]{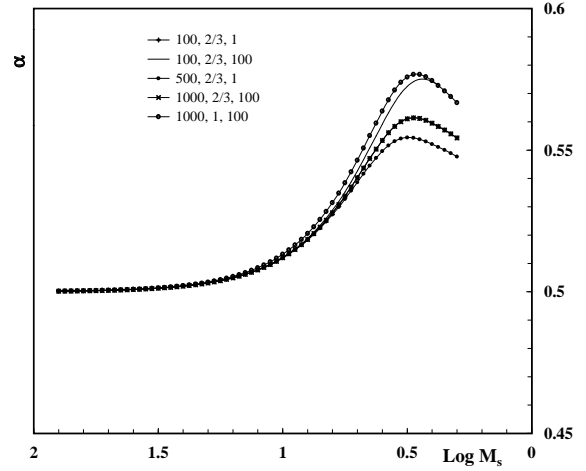}
\caption{The evolution of the integral radio spectral index with shock Mach number. Curves are marked by the values of parameters  ${\cal M}_{\rm 0} ,k ,c_{\rm 1}$}
\label{fig5}
\end{figure}

The radio spectral index is the most important parameter of SNRs. Our model gives very specific predictions for this parameter. The evolution of the integral spectral index with ${\cal M}_{\rm s}$  is plotted in Fig.\ref{fig5}. The value of the spectral index is calculated using the formula:
$\alpha  = \int\limits_{r_{in} }^1 {\alpha (r) \cdot \varepsilon_\nu (r)r^2 dr} /\int\limits_{r_{in} }^1 {\varepsilon_\nu (r)r^2 dr}$ for any given ${\cal M}_{\rm s}$.  First of all, as can be seen in Fig.\ref{fig5}, the shape of all curves is very similar; the spectral index increases with decreasing ${\cal M}_{\rm s}$ (reaching the value of $ \la 0.6$ at ${\cal M}_{\rm s}  \le 4$)  until it achieves its maximum around ${\cal M}_{\rm s}\cong 3$ and then slowly decreases. Although the shape of all curves are similar, very little depending on the input parameters, the maximum of the spectral index $\alpha_{\rm max}$ depends on the input parameters of the model, namely, $\alpha_{\rm max}$ is larger for small ${\cal M}_{\rm 0}$  at fixed other input parameters and it increases with increasing $c_{\rm 1}$. If we take the quadratic relationship between $c_{\rm 1}$  and ${\cal M}_{\rm 0}$ into account, the last conclusion implies that $\alpha_{\rm max}$ is higher for high values of the magnetic field strength.

\emph{The general properties of the model that have practical importance are the shapes of}  $\sigma_\nu=\sigma_\nu ({\cal M}_{\rm s} )$ and $\alpha=\alpha({\cal M}_{\rm  s})$ \emph{dependencies and the fact that they have a universal nature in the sense that they barely depend on the poorly known injection parameters.}

 %%%%%%%%%%%%%%%%%%%%%%%%%%%%%%%%%%%%%%%%%%%%%%%%%%%%%%%%%%%%%%%%   
\section {Application of the model to the real SNRs. $\Sigma-D$ relationship}
 In the study of SNRs the empirical $\Sigma - D$ dependencies play an important role. This relation is used as the main tool for testing the results of the theory. Moreover, many statistical studies of Galactic SNRs have relied on the $\Sigma - D$  relations to derive distances for individual SNRs. % from their observed flux densities and angular sizes. %It is difficult to use the results In this section we 
The results of the previous section are presented in a functional form with arbitrary normalization and, also, without taking into account the possible onset of the radiative phase in the course of SNR evolution. In this section we apply these results to the real SNRs evolving in real interstellar environments.
\subsection{Input parameters}

To apply our model to the real SNRs, we first have to specify additional new parameters concerning the SNR itself (the ejecta mass $M_{\rm ej}$, the initial kinetic energy of the ejecta $E_{\rm SN}$, the rms velocity of the ejecta ${\rm v}_{\rm 0}$), and the ISM (the total pressure $P_{\rm 0}$, the density of plasma $n_{\rm e0}$, the magnetic field strength $B_{\rm 0}$).

 The beginning of the model is parameterized by the initial radius $R_{\rm 0} $ and Mach number ${\cal M}_{\rm 0} $ of  the shock wave at which the Sedov phase begins. We ignore the ejecta stage because the possible contribution of this part of the remnant to the total emission with time will rapidly decrease.  

 In the following we assume a fully ionized gas with a ratio of specific heats $\gamma  = 5/3 $ and with a helium abundance relative to hydrogen as $n_{\rm He}= 0.1\,n_{\rm H}$.  The number density  and electron density are $n=n_{\rm He}+n_{\rm H}=1.1\,n_{\rm H}$  and  $n_{\rm e}= n_{\rm H} + 2\,n_{\rm He}=1.2\,n_{\rm H}$, respectively. The mean mass per particle and per electron in units of the proton mass, $m_{\rm p}$, are $\bar \mu  = (1.4/2.3)$  and $ \mu  _{\rm e}= (1.4/1.2)$, the total number of particles per electron is $ 2.3/1.2 \approx 1.92$. The mass density of matter we express in terms of the electron concentration $n_{\rm e}$: $\rho  = \mu _{\rm e} m_{\rm p} n_{\rm e}$.

To take very extended SNRs into consideration, we use an analytical approximation of Cox \& Anderson (\cite{coxa82}). The details of this approximation are given in Appendix 1. The pressure of ISM during the late stages  becomes a leading factor in defining the evolution of the SNR, so it is important to include the pressure of the magnetic fields $P_{\rm M} = B^2 /8\pi$ in the total pressure of matter in the ambient ISM and take  the effect of the magnetic field into account in determining the value of the compression ratio at the shock front. For simplicity we make no difference between the thermal gas and cosmic ray components when we determine the total pressure in the ambient medium, and we assume that there is an equipartition between the three components of the ISM. 

Since the maximum speed of propagation of small perturbations in the magnetized medium is the speed of magneto-sound waves, therefore, as a measure of intensity of the shock wave, we use the ratio of the velocity of the shock wave ${\rm v}_{\rm s}$ to the speed of fast magneto-sound waves $c_{{\rm MS}}  = (c_{\rm s}^2  + {\rm v}_{\rm A}^2 )^{1/2}$  (where ${\rm v}_{\rm A}$  is the Alfven speed, $c_{\rm s}$ the sound speed) in ISM : ${\cal M} = {\rm v}_{\rm s} /c_{{\rm MS}} $ and for simplicity, we refer to this value as the shock Much number. 
% Numerically for $c_{\rm MS}$  we have ($\gamma  = 5/3$) $c_{{\rm MS7}}  = 2.02\,\,(5\beta /6 + 1)^{1/2} \,\,H_{\rm  - 5}\,\,n_{\rm e2}^{\rm -1/2}$  where $c_{\rm MS7}$  is $c_{\rm MS}$  in  $10^7 {\rm cm/s}$ and $n_{\rm e2} = 10^{ - 2} n_{\rm e0}$.
Using the shock jump conditions for the perpendicular plane-parallel shock, we can find the compression ratio  as the real root of the quadratic equation (Field et al. \cite{field}):
   \begin{displaymath}
\frac{{2(2 - \gamma )}}{{2 + \gamma \beta }}\,\frac{x_{\rm s}^2  }{{{\cal M}^2 }}+ \left( {\gamma  - 1 + \frac{{2\gamma (\beta  + 1)}}{{2 + \gamma \beta }} \frac{1}{{{\cal M}^2 }}} \right)x_{\rm s}  - \left( {\gamma  + 1} \right) = 0,
   \end{displaymath}
where $\beta  = P_{{\rm th}} /P_{\rm M}$   ($P_{\rm th}$ is the sum of the thermal gas and cosmic ray  pressures) - one of the main input parameters characterizing the ISM in our model. This equation provides a necessary condition that $x_{\rm s}  \to 1$  with ${\cal M}  \to 1$; and when $\beta  \to \infty $  (${\cal M}$ is ordinary sonic Mach number), it gives for the compression ratio the well-known relation (Landau \& Lifshitz \cite{ll86}): 
   \begin{displaymath}
x_{\rm s} ({\cal M}) = \frac{{\gamma  + 1}}{{\gamma  - 1 + 2{\cal M}^{ - 2} }}\,\,.
   \end{displaymath}
The total pressure of the ISM can be expressed in terms of the interstellar magnetic field strength and the parameter $\beta$ 
\begin{equation}
P_{{\rm 04}}  = 2.88 \, (\beta  + 1)B_{{\rm  - 5}}^2 
\label{eq29}
\end{equation}
where $P_{\rm 04}$  is $P_{\rm 0}$  in units of  $10^4 \,\,{\rm K} \,{\rm cm}^{ - 3}$.
For the initial Mach number we now obtain 
\begin{equation}
{\cal M}_{\rm 0}  = 458.1\left( {\frac{{\mu_{\rm e}\,n_{{\rm e0}} }}{{5\beta/6  + 1}}} \right)^{1/2} \frac{{{\rm v}_{{\rm 09}} }}{{B_{{\rm  - 5}} }}
\label{eq30}
\end{equation}
where %
 ${\rm v}_{09}$  is initial shock velocity in $10^9 \,{\rm cm/s}$.
  The mass of the material ejected at the SN explosion, $M_{\rm ej}$,  and its kinetic energy,  $E_{\rm SN}$, are used as input  parameters of the model characterizing the supernova event. The relation between them and ${\rm v}_{09}$ is  
\begin{equation}
m_{{\rm ej}}  = E_{051} /{\rm v}_{09}^2 
\label{eq31}
\end{equation}
where $E_{\rm 051}$ is $E_{\rm SN}$  in $10^{51}\,\,{\rm erg}$, and $m_{\rm ej}$  is $M_{\rm ej}$ in solar masses.

A very important parameter of the model is the initial shock radius $R_{\rm 0}$ at which the Sedov phase begins. Usually the beginning of the Sedov phase is determined from the condition that swept-up mass becomes equal to the ejected mass, e.g., at $\chi  \cong 1$ in $4\pi R_0^3 \rho _0 /3 = \chi m_{\rm ej}$.  As the transition from the free-expansion phase to the Sedov phase is smooth and occurs over a finite period of time, it is difficult to choose the concrete value for $R_{\rm 0}$. Moreover, as shown by Cioffi et al. (\cite{cioffi}) at some conditions (not far from the typical ones for real SNRs), the interior of the adiabatic blast wave may not settle completely to the Sedov similarity solution.  We determine the value of $R_{\rm 0}$  from the law of motion  (A1) and its new representation  (A2); namely, equating them we have $(36/5)R_{\rm c}^3  = R_0^3 \,{\cal M}_0^2$   from which taking into account the definition of the initial Mach number as  $ {\cal M}_0  \equiv {\rm v}_{\rm 0} /\left( {\gamma P_0 /\rho _0 } \right)^{1/2} $ and (31), we obtain %  
\begin{equation}
R_{{\rm 0pc}}  = 1.87 \cdot ( m_{{\rm ej}} /\mu_{{\rm e}}n_{{\rm e0}} )^{1/3} 
\label{eq32}
\end{equation}
where $R_{\rm 0pc}$  is $R_{\rm 0}$  in pc. The comparison of this formula with the one   derived by introducing the parameter $\chi$, $R_{{\rm 0pc}}  = 2.13 \cdot (\chi m_{{\rm ej}} /\mu_{{\rm e}}n_{{\rm e0}} )^{1/3} $, gives $\chi\approx 0.6$.
    In our case the low values of $\chi$  are physically acceptable because the self-similar structure in the shock interior begins to form earlier in the free expansion phase. Initially it occupies a small part of the volume behind the shock front, but with time the fraction of the volume with self-similar structure will increase. As mentioned above, we consider only the volume between $R_{\rm in}$ and $R_{\rm s}$, where the Sedov structure holds, and ignore the emission from the inner volume of the remnant, which is occupied with ejecta material. 

When the downstream temperature approaches some ``cooling temperature'' $T_{\rm c} \sim 10^{\rm 6}$~\rm{K}, the remnant will make a transition from the Sedov regime to the radiative or ``pressure-driven snowplow'' phase (McKee \& Ostriker  \cite{mo77}; Cioffi et al. \cite{cioffi}). A cold dense shell begins to form when the first portion of the shocked gas has radiated most of its thermal energy.   Since we consider the acceleration of electrons from the high-energy tail of the downstream Maxwellian distribution function, it is natural to assume that injection of fresh particles into the process of acceleration stops when radiative cooling of the shocked gas first starts.
The velocity at the beginning of the radiative phase according to Cioffi et al., (\cite{cioffi}) is
   \begin{displaymath}
{\rm v}_{{\rm PDS}}  = 413 \,  E_{{\rm 051}}^{1/14}  \, n_{{\rm 0}}^{1/7} \,\, {\rm km/s}.
   \end {displaymath}
The process of the radiative cooling was studied by many authors, and their results did not always agree. For instance,  the same functional formula is derived by Franco et al. (1994), but with a lower value of the coefficient of $265$.  In the following we employ the formula of Cioffi et al. (\cite{cioffi}) because the authors of this work have given a more precise description of the  preceding evolution of the shock wave and considered the very beginning of shell formation.
In our notations the above expression takes the form
 \begin{equation}
{\rm v}_{{\rm sf7}}  = 4.02\,\,E_{{\rm 051}}^{1/14}  \, n_{{\rm e0}}^{1/7}. 
\label{eq33}
\end{equation}

The shock wave Mach number at which thin shell formation begins is
\begin{equation}
{\rm  }{\cal M}_{{\rm cool}}  = 18.42\,\left( {{{5\beta }}/{6} + 1} \right)^{ - 1/2}\,\mu_{e}^{1/2}\,\,E_{051}^{1/14} \,n_{{\rm e0}}^{9/14} \, \,B_{{\rm  - 5}}^{ - 1}. 
\label{eq34}
\end{equation}
From this formula we can find the density of the ISM at which SNR will evolve without shell formation up to some critical value of   ${\cal M}_{\rm cr}$ as
\begin{equation}
n_{{\rm e0}} \la
1.08 \times 10^{ - 2} \,\,\frac{{(5\beta /6 + 1)^{7/9} B_{ - 5}^{14/9} {\cal M}_{cr}^{14/9} }}{{\mu _e^{7/9} \,E_{051}^{1/9} }}.
\label{eq35}
\end{equation}
As the pressure in the ISM is the quantity that is to a lesser degree subjected to strong fluctuating, it is useful to rewrite this expression in terms of this quantity. Using Eq.~\ref{eq29} we have
\begin{equation}
n_{{\rm e0}}  \le 4.74\times10^{-3}\,\left( {\frac{{5\beta /6 + 1}}
{{\beta  + 1}}} \right)^{7/9} \frac{{P_{04}^{7/9} {\cal M}_{cr}^{14/9} }}{{\mu _e^{7/9} \,E_{051}^{1/9} }}.  %
\label{eq36}
\end{equation}

The SNR with  $E_{051}  \approx 1$ evolving in the ISM with  $\beta  \approx 2$ (equipartition between thermal, CR and magnetic pressures), $\mu_{\rm e}\approx 1.4/1.2$, and $P_{04}\approx 3$ (Boulares \& Cox ~\cite{bcox}) will finish its life by merging with the ISM (${\cal M}_{cr}\approx2$) before cooling becomes important, if  $n_{\rm {e0}}  \le  2.65\times 10^{-2}\,\, \rm {cm}^{ - \rm{3}} $. As shown above, the break in the $\Sigma-D$  track to a steeper power law starts at ${\cal M}_{\rm break} \sim 10 $ and, for this to happen without dense shell formation, the density of the ISM must satisfy the condition $n_{\rm {e0}}  \le  0.23\,\, \rm {cm}^{ - \rm{3}}$. The constraint on the density is not very strong, so, for many remnants in the course of their evolution the break in the  $\Sigma-D$ track can occur prior to the beginning of the radiative phase.  

As a result, if the density of the ISM satisfies the value given in Eqs.~(\ref{eq35}) and (\ref{eq36}), then the break in  the $\Sigma-D$ evolutionary track occurs due to the steeper spectrum of the newly accelerated electrons, which happens at a diameter 
 \begin{equation}
D_{\rm{break}}  \approx 80.9\,(E_{051} /\,P_{04} )^{1/3} \,\,\,(\rm{pc}).
\label{eq37}
\end{equation}
In the opposite case, when SNR evolves in higher density ISM, the brake in the $\Sigma-D$ track occurs due to the onset of the radiative cooling at
\begin{equation}
D_{\rm{break}}  \approx 36.4\,(\chi /\mu _{\rm e} )^{1/3}\, E_{051}^{2/7} \,n_{\rm{e0}}^{\rm { - 3/7}} \,\,\,(\rm{pc}).
\label{eq38}
\end{equation}
In deriving Eqs. \ref{eq37} and \ref{eq38}, we have assumed  ${\cal M}_{\rm {break}}=10$, $\beta  = 2$. Note that in the case of radiative cooling, the track has sharp break to steeper power law but, in the case of DSA, the track steepens gradually, and it is difficult to determine the precise diameter $D_{\rm break}$ and Eq.~\ref{eq37} determines the diameter of the remnant at which the surface brightness just begins to break. %

Finally, two additional parameters concerning the mechanism of acceleration $\eta$ and  $\psi$ remain as free parameters of the model. %As was shown above though the dependence of the model results on $\psi$ is weak the dependence on $\eta$ is linear.
In principle, special radio, X-, and $\gamma$-ray observations of an individual remnant can be used to estimate these parameters. Here we aim to examine  the capability of DSA to explain the averaged radio properties of shell-type SNRs and to reveal the principal manifestations of the action of DSA in SNRs. As shown in the following section, the statistical $\Sigma  - D$  relation can be used to obtain an order-of-magnitude estimation for these parameters. In model calculations for real SNRs, we have used an estimation of $\eta  = 6 \times 10^{ - 4}$, which is derived there.

%xxxxxxxxxxxxxxxxxxxxxxxxxxxxxxxxxxxxxxxxxxxxxxxxxxxxxxxxxxxxx
\subsection{ The $\Sigma  - D$ relationship}
\begin{figure*}
%\centering
\sidecaption
\includegraphics[width=12cm] {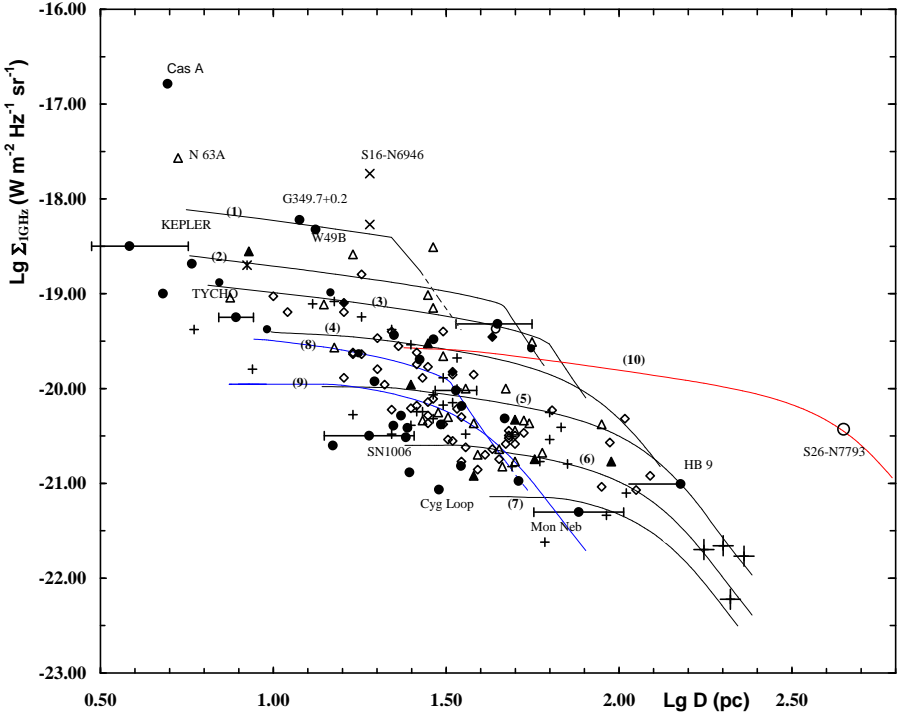} %
%\resizebox{\hsize}{!}{\includegraphics{SD.eps}}
%
\caption { %  
   ~$ \Sigma_{\rm 1GHz}  - D$  diagram for 158 composite and shell-type SNRs and SNR candidates in our Galaxy (filled circles), M31 (small crosses), M33 (open diamonds), LMC (open triangles), SMC (filled triangles), NGC 300 (filled diamond), NGC 6946 ($\times$ 's), NGC 7793 (open circles), and IC 1613 (asterisk). Galactic loops are shown as large crosses and the only candidate for HNR as a large open circle. % 
 For several galactic SNRs, error bars due to the distance uncertainty are shown, and for one  galactic SNR (HB 9) only upper limit for the distance is known. Theoretical  $\Sigma  - D$ tracks are labeled according to the numeration % 
in Table~\ref{tab2}.}
\label{fig6}
\end{figure*}

Empirical $\Sigma  - D$ relations are very useful tools for testing various theoretical models of SNR evolution.  A brief review of this problem is given in the recent work by Urosevich et al. (\cite {uros2005}). In Fig.~\ref{fig6} the surface brightness at the 1 GHz($\Sigma_{\rm 1GHz}$) -- diameter diagram is presented, which consists of 158 large-diameter %
SNRs and candidates for SNRs with known distances: 33 composite and shell-type   Galactic SNRs, 4 Galactic Loops, 25 in LMC, 7 in SMC, 
51 in M33, 30 in M31, 3 in NGC 300, 2 in NGC 6946, 2 in NGC 7793, and one SNR candidate in IC 1613.
 For the galactic SNRs, we used only the data given in the catalogue of Green (\cite{green,green1})  with one exception -- we did not include the SNR OA184 in our consideration because, as shown in the recent work by Foster et al. (\cite{foster}), this object is most probably an HII region, not an SNR. %with one exception - it is excluded the SNR OA184, as it is shown in recent work by Foster et al. (\cite{foster}) this object very probably is H II region.} %
The data on the Galactic loops  are  from Berkhuijsen (\cite{ber86}). 
 The data for the remaining 121 SNRs in nearby normal galaxies are from the work of Urosevich et al. (\cite {uros2005}). We did not use the remnants in the sturburst galaxies because, on the one hand, the majority of them are very young ones and, on the other, the physical conditions in the ISM of these galaxies differ from the standard conditions in the normal galaxies considered in our model. %

In the sample of SNRs presented in Fig.\ref{fig6} we also did not include Radio supernovae and plerionic SNRs because the mechanism of their emission  differs radically from that of our model. Our model also cannot explain the radio emission from the young SNRs actively interacting with the dense circumstellar structures or else in the double-shock stage of development.  For instance, the radio emission from the Cas A SNR is commonly considered as the result of the interaction of the SN ejecta with circumstellar material. The radio emission from other young  historical SNRs Tycho, Kepler, and AD1006, in principle, can be explained in the framework of our model. 
 Further, our model meets with difficulty in explaining  the radio emission from SNRs with power-law spectra that are too flat or too steep.%

Some evolutionary tracks are also shown in Fig.~\ref{fig6} and the parameters of the models used to calculate them are given in Table 2. In all runs, we used an estimate of $ 6 \times 10^{-4}$ for the  injection parameter $\eta$, derived as follows. For each value of  surface brightness in the $\Sigma_{\rm 1GHz}- D$ diagram, the largest observed diameter is considered to be the value of $D_{\rm break}$ due to the radiative cooling (Eq.~\ref{eq38}). The largest diameter remnants for a given surface brightness occupy the upper right region in the empirical $\Sigma_{\rm 1GHz}- D$ distribution, which should be reasonably free of selection effects. For reliable estimates, it is desirable to consider remnants with intermediate values of $\Sigma$ because the brightest, largest SNRs can be the ones for which our mechanism is not applicable and, in the opposite case, the largest SNRs with low $\Sigma$ are affected by selection effects. As seen in Fig.~\ref{fig6}, Galactic SNR Kes 67 (G18.8+0.3) with $\lg \Sigma _{{\rm 1GHz}} = -19.58$ and $\lg D ({\rm pc}) = 1.75\ $  satisfies the selection requirements. As a turning point of the track, along which this remnant evolves, we take the point with coordinates $\lg \Sigma _{{\rm 1GHz}} = -19.55$ and $\lg D ({\rm pc}) = 1.75$ because, besides Kes 67, one more remnant, B0547-697 in LMC with $\lg \Sigma _{{\rm 1GHz}} = -19.51$ and $\lg D ({\rm pc}) = 1.75\ $, is located around this point. %
Using these data in  Eq.~\ref{eq38} we  find $n_{\rm e0}=.32$ $\rm cm^{-3}$. Then, using the latter  value in Eq.~\ref{eq27} and adopting $B_{-5}\approx 0.5$ and $E_{051}= 1$, we get an estimation of  $\eta\approx 7.0\times10^{-4}$. Varying the values  $B_{-5}$ and $E_{051}$ within reasonable limits $0.3 \div 1.0$ and $0.5 \div 2$, respectively, we find a range for $\eta$ of $ (3\div 11)\times10^{-4}$. In the calculations we have used  $\eta = 6\times10^{-4}$. 
\begin{table}
\caption{Input parameters of the model runs. $E_{\rm 051}$ is $E_{\rm SN}$ in units of $10^{51}\,\rm {erg}$,  $B_{\rm -5}$ is $B_{0}$ in units of $10^{-5}\,\rm {G}$, $D_{break}$ is in $\rm{pc}$. In all rans $k=2/3$, $\eta = 6\times10^{-4}$, $\beta=2$, $m_{\rm {ej}}=1.4$ }
\label{tab2}
%\medskip
\centering\begin{minipage}{8cm}
\begin{tabular}{rrrrrrr} \hline
 \hline 
No& $n_{\rm e0}$ & $E_{\rm 051}$ & $B_{ -5}$ & $D_{break}$ &  ${\cal M}_{cool}$ & $c_{1}$\\ 

\hline
   1 & 5.0 & 1 & 0.6 & 22 & 37& 5.40 \\ 
   2 & 1.0 & 1 & 0.6 & 46 & 13& 5.40 \\ 
   3 & .5 & 1 & 0.5 & 60 & 10& 4.50 \\
   4 & .1 & 1 & 0.5 & 63 & 3&4.50  \\ 
   5 & .05 & 1 & 0.3 & 88 & 4& 2.70 \\ 
   6 & .005 & 1 & 0.3 & 89 & 1&2.70  \\ 
   7 & .001 & 1 & 0.3 & 88 & 1& 2.70\\
   8 & .5 & 0.1 & 0.5 & 29 & 9& 0.45\\
   9 & .1 & 0.1 & 0.5 & 29 & 3& 0.45\\
   10& .01 & 50 & 0.3 & 323 & 2& 135.0\\
\hline
\end{tabular}
\end{minipage}
\end{table}

First of all, it is important to note that the shapes of the model's evolutionary tracks are in very good accordance with the claim made by Berkhuijsen (\cite{ber86}) that SNRs may evolve at nearly constant $\Sigma$ followed by a steep decrease. In our model such behavior of the $\Sigma- D$ tracks has a natural explanation. In the low-density ISM, when SNR evolves without the onset of the radiative cooling, this is the direct consequence of the action of DSA, but when SNR evolves in the high density ISM, it follows from the assumption that acceleration of electrons takes place from the downstream thermal distribution. This assumption also naturally results in the tight correlation between the surface brightnesses in radio and X-ray wavelengths revealed by Berkhuijsen (\cite{ber86}).

From the shapes of the evolutionary tracks and the very wide possible range of variation in the surface brightness for a given diameter (up to 3 orders of magnitude), it follows that there is no physically meaningful $\Sigma - D$ relation that can be used as the tool for estimating the distances to SNRs. For example, two identical SNRs evolving in identical environments but with different diameters (and ages) may have very close values of $\Sigma$. On the other hand, for two SNRs with the same $D$ the values of $\Sigma$ may differ by several  orders of magnitude.  

To see this, let us write the expression for $\Sigma$ (Eq.~\ref{eq27}) as  

\begin{equation}
\Sigma _{\rm 1GHz}  \propto n_{\rm e0}^{2/3} B_{\rm 0}^{1.5} E_{\rm SN}^{1/2}M_{\rm ej}^{-1/6} \propto n_{\rm e0}^{2/3+1.5k_{0}} E_{\rm SN}^{1/2}\,M_{\rm ej}^{-1/6},
\label{eq39}
\end{equation}
where it is assumed that in the ISM the magnetic field is correlated with the density as $B_{0}  \propto n_{{\rm 0}}^{k_{0}}$, which can easily give a range of variation in $\Sigma$ up to 3 orders of magnitude. Note that in the work of Berezhko \& V\"olk (\cite {berezhko}), the value of $\Sigma_{\rm R}$ at a given diameter depends only on the SN energy as $E_{\rm SN}^{7/4}$.  

Concerning the ``maximum observable diameter'' discussed by Berkhuijsen (\cite {ber86}), our model gives the relation 
\begin{equation}
\Sigma_{\rm 1GHz} \propto E_{\rm SN}^{17/18 + k_{0}} \,D_{\rm max }^{ - (14/9 + 7k_{0} /2)};
\label{eq40}
\end{equation}
in particular, at $k_{0} \sim 0.5$ we have the relation $\Sigma  \propto D_{\rm max }^{ -3.5}$ derived in (Berkhuijsen \cite {ber86}).

\begin{figure}
\centering
%\resizebox{\hsize}{!}{\includegraphics{fig7F.eps}}
\includegraphics[width=7.5cm]{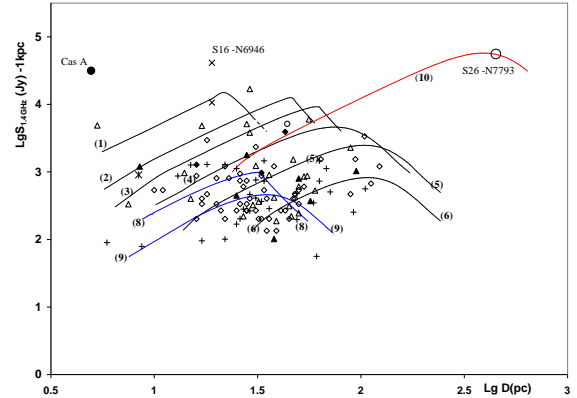} %
\caption{1.4 GHz flux density - diameter relation for 121 SNRs in nearby normal galaxies + Cas A SNR, if they are at $1~\rm {kpc}$ distance. Symbols and lines have the same meanings as in Fig.~\ref{fig6}.}
\label{fig7}
\end{figure}

The observational $\Sigma - D$ relations, usually presented in the power-law form ($\Sigma \propto D^\beta$), are mainly due to: 1) presence of SNRs with diameters from a wide range of sizes from small-diameter bright young SNRs and large faint old SNRs; 2) $D^{-2}$ effect; 3) selection effects. Unfortunately, it is difficult to check the influence of various selection effects on this problem. For instance, the absence of a statistically meaningful
$\Sigma - D$ relation for the sample of SNRs in M33, which is particularly incomplete in the smallest and extended faint remnants, was emphasized by Gordon et al. (\cite{gordon}).
In Fig.~\ref{fig7} the 1.4 GHz flux density - diameter relation is presented for 121 SNRs in nearby normal galaxies if they are placed at 1 kpc distance. All the data are from Urosevich et al. (\cite {uros2005}). For an equidistant sample of remnants, this relation can be used as an equivalent to the $\Sigma - D $ relation, but it is free of the $D^{-2}$ effect. As noted above (see Sect.~2.3) the effect of $D^{-2}$ results in a smaller difference between the different $\Sigma - D$ tracks than it is for luminosity - diameter tracks with the same set of the input parameters.   

Thus, in the formation of $\Sigma- D$ relation, the two main parameters $n_{\rm 0}$ and $E_{\rm SN}$ have equal importance (See Eq.~\ref{eq39}). At intermediate diameters, the contribution of $n_{\rm 0}$ is more significant, but $E_{\rm SN}$ critically determines the maximum size of the radio remnant, especially, in low-density environments. As can be seen from Fig.~\ref{fig6}, practically all the remnants can be described within the range  of $(10^{\rm 50} \div 10^{\rm 51})~{\rm erg}$. Varying mainly the value of  $E_{\rm SN}$, the model is able to explain the existence of very large diameter SNRs and even the remnants of extremely energetic SNe, known as hypernovae (HNRs). For instance, our model easily explains the radio emission from the Galactic loops, so 
there is no need for energies higher %
than the standard value of $ \sim 10^{51}\, {\rm erg}$. On the contrary, according to the model for the existence of such a large radio source as the SNR candidate N7793-S26 in the galaxy NGC7793,   at least $ \sim  5 \times 10^{52}\, {\rm erg}$ energy is required. Although in Fig.~\ref{fig6} only one track passing through this object is shown, of course, this is not the only possible curve; but for any plausible set of other input parameters, this constraint on energy is indispensable for explaining the radio emission from this object.  Such energies are characteristic of hypothetical Hypernova explosions, and this result can be considered as evidence of the real existence of such objects, and the SNR candidate N7793-S26  can be regarded as a real HNR. It is important to note that no evidence has been found  of an interior star cluster, the detection of which would argue in favor of the collective SNR hypothesis (Pannuti et al. \cite{pann2002}). The idea that the detection of the very large diameter radio remnants may be considered as proof of the real existence of hypernova was proposed in our early work (Asvarov \cite{asv2000}), where the parameters of expected hypernovae remnants were calculated.
  As seen in Fig.~\ref{fig6} at intermediate diameters ($\la 100 \rm pc$), the adiabatic radio remnants of Hypernovae, evolving in the ISM with typical parameters, do not differ in essential features from the ordinary SNRs. Of course, in the early stages of their evolution the hypernovae can be very bright in radio, depending on the properties of the circumstellar matter, % 
but at the same time, the high luminosity  of a very young SNR does not mean that we are necessarily dealing with a hypernova.
  
We must note that in terms of other known models, e.g., the model of van der Laan (\cite{laan}) and its modifications, describing the radio evolution of evolved shell-type SNRs, it is not easy to explain the existence of SNRs with such large diameters

Within the framework of the present model, a number of features of the empirical $\Sigma  - D$ relation has a simple explanation. For instance, the small number of remnants with small diameters and low $\Sigma$ (lower left corner in the diagram) is the result of the very fast evolution of the SN blast wave in the low-density ISM where SNRs have low $\Sigma$. According to our model, it is easy to account for a high concentration of SNRs at diameters of $20-50 ~\rm pc$ in the $\Sigma  - D$ diagram: various evolutionary tracks intersect at these diameters and the sample of remnants here consists of objects evolving at different initial values of $E_{\rm SN}$ and $n_{0e}$. However, it seems more likely that this excess comes from the energies of most SNRs being closer to $10^{50}~\rm erg$ than $10^{51}~\rm erg$ (Lozinskaya  \cite{lozi}). Another fact in favor of such a possibility is the absence of SNRs near the break points of $10^{51}~\rm erg$ tracks in a high density medium (curves (1) - (3) in Fig.~\ref{fig6}), although this part of the $\Sigma  - D$ distribution is free of the selection effects. Unfortunately, the importance of selection effects at large diameters is hardly known, but if most SNRs have initial energies of $~\sim 10^{\rm 51}~{\rm erg}$, then it follows from the comparison of the model with the $\Sigma  - D$ distribution that there should be a large population of $150 - 250$ pc diameter SNRs with $ \Sigma _{{\rm 1GHz}}  \le \,\,10^{ - 22} \,\,{\rm W}\,{\rm m}^{{\rm  - 2}} \,{\rm sr}^{{\rm  - 1}} \,{\rm Hz}^{{\rm  - 1}}$. Identification of such low-brightness remnants in our Galaxy is a very difficult task, although there are a few successful attempts (Reich \cite {reich}); but they can be detected in nearby galaxies.

From the distribution of the model tracks in the $\Sigma  - D$ diagram,  a very important conclusion can be made that shell-type SNRs with $ \Sigma _{{\rm 1GHz}}  \le \,\,10^{ - 20} \,\,{\rm W}\,{\rm m}^{{\rm  - 2}} \,{\rm sr}^{{\rm  - 1}} \,{\rm Hz}^{{\rm  - 1}}$ are located in the low-density phase of the ISM. 
It is important to take into account the fact  that our sample of Galactic SNRs consists of remnants for which the distances are mainly determined   from the interaction of the supernova  blast wave with the dense environments; i.e. our sample consists of objects located in special environments, e.g., near large atomic or molecular clouds. Indeed, for $40$ Galactic SNRs with estimated distances out of 101 SNRs (without plerions), for which  the $ \Sigma _{1{\rm GHz}}$ can be calculated in the catalogue of Green (\cite{green}), the mean value of the surface brightnesses $\Sigma _{1{\rm GHz}}$ (in standard units) is $4.82 \times 10^{ - 19}$ (\- $7.47 \times 10^{ - 20}$ without Cas A), but the  mean value of $\Sigma _{1{\rm GHz}}$ for remaining 61 SNRs (without distance estimations) is  $7.45 \times 10^{ - 21}$.

Finally, the fact that 63 (62\%) out of those 101 SNRs have $ \Sigma _{{\rm 1GHz}} <  \,\,10^{ - 20} \,\,{\rm W}\,{\rm m}^{{\rm  - 2}} \,{\rm sr}^{{\rm  - 1}} \,{\rm Hz}^{{\rm  - 1}}$ can be considered as a proof that the large fraction of SNRs are evolving in the low-density phase of the ISM.
 The same conclusion about the environments of the galactic SNRs  was earlier made by  Marsden et al. (\cite {marsden}) from the association of AXP and SGR with SNRs and in Koo and Kang (\cite {koo}) from the considerable deficit of $\rm {HI}$-emitting, old SNRs. This result is also in accordance with the theory of McKee and Ostriker (\cite{mo77}), which predicts the widespread existence of hot and low-density gas in the Galaxy.

%Hot gas blown from within the remnant results in formation{education} of a new shock wave which surrounds a dense shell. It is obvious, that the front of this new wave will not be a source of a radio emission because for acceleration of new particles on this wave there are no necessary conditions.

%%%%%%%%%%%%%%%%%%%%%%%%%%%%%%%%%%%%%%%%%%%%%%%%%%%%%%%%%%%%%%%%%%
\subsection{Spectral index}
The statistics of spectral indices is another argument favoring the DSA mechanism. The catalogue of SNRs of Green (\cite{green}) contains 231 SNRs from which  180 $(78\%)$ are  shell-type (labeled in the catalogue ``s'' and ``s?'') for 91 of which the values of the spectral indices are determined: for 87 SNRs the mean value of the spectral index is  $\bar \alpha  = 0.506 \pm 0.111$ and 4 SNRs have variable spectral indices.  Twenty-one of these 91 SNRs $(23\%)$ (7 SNRs with $\alpha  > 0.65$, 10 SNRs with $\alpha  < 0.4$ and  4 SNRs with variable $\alpha$) have spectral indices that are not in agreement with DSA: our model predicts $\alpha$ in the  range of (0.5 - 0.6). The occurrence of shell-type SNRs with a flat spectrum can be explained by the inclusion of second-order Fermi acceleration (Schlickeiser \& Furst \cite{sf89}; Ostrowski \cite{ost99}) or by the effect of compression of the interstellar cosmic ray electrons by the radiative shocks (Cox et al. \cite{coxshelton99}). Indeed, there is some evidence that flat spectra are characteristic of SNRs interacting with higher density environments.

 Steeper spectrum SNRs are as a rule the young ones (for example, the historical shell remnants Cas A, Tycho, Kapler, SN 1006  all have spectral indices steeper than $\alpha = 0.6$), which can be explained by the nonlinear modification of the shock structure (see, for the recent review, V\"olk \cite{volk2006}),   but in the case of evolved remnants there is no more or less satisfactory explanation for such $\alpha$.
In the framework of the DSA mechanism in the test particle approximation,  the values of $\alpha$ steeper than $ 0.6$ can be explained if we assume that the remnant evolves in a pre-existing cavity with a radially increasing density distribution.   

% Green argued that it is difficult, however, to determine the index to better than 0.1

Our model does not predict any detectable dependence of the spectral index on the diameter because the universal character of the $\alpha({\cal M}_{\rm s})$-dependence (Fig.\ref{fig5}) does not lead to the same character for the $\alpha(D)$ - dependence. But, in any case,  the model predicts spectral indices approaching the value of $\sim 0.6$ for large diameter SNRs.
According to our model, the SNRs with Mach numbers ${\cal M}_{\rm s}  \le 5$ will have the mean value of $\alpha  \approx 0.55 \div 0.60$. Assuming for simplicity that SNR evolves according to the Sedov law, ${\cal M}_{\rm s}  \propto t^{ - 3/5}$,  the number of SNRs with Mach number smaller than ${\cal M}_{\rm s}$ is $N( \le {\cal M}_{\rm s} ) \propto {\cal M}_{\rm s}^{ - 5/3}$,  from which it follows that the number of SNRs with $\alpha  \ge 0.55,\,\,\,N(\alpha  \ge 0.55) = N({\cal M}_{\rm s}  \le 5)$ must be about $70\%$ of the total number of SNRs. Here we have adopted ${\cal M}_{\rm f}=2.5$ for the final Mach number  at which $\Sigma$ drops more than two orders of its initial value, and the SNR becomes invisible.
From 92 SNRs only $18 (20\%)$ (excluding 3 historical young SNRs: SN1006, Kepler, Tycho) have $0.55 \le \alpha  < 0.65$. The shortage of SNRs with such spectral indices can be easily explained by various selection effects, all of which are biased against the faint large-diameter SNRs.

\subsection{ Radiative SNRs}

 Since electrons, by assumption,  are injected from the high-energy tail of the downstream Maxwellian distribution function, the process of generating new radio-emitting electrons stops with the beginning of the radiative cooling. This implies that, with the beginning of the radiative phase, the radio emission from SNR starts to drop rapidly. On the other hand, with the onset of radiative cooling the emission from an SNR in X-rays will also decrease rapidly. Indeed, a tight correlation between radio and X-ray surface brightnesses has been revealed by Berkhuijsen (\cite {ber86}),  which confirms our suggestion about the injection of electrons into the DSA. 

  As a consequence of this, our model predicts the absence of radio emission from the radiative remnants evolving in the  homogeneous ISM according to the standard theory of SNR evolution and the existence of radio-quiet radiative SNRs emitting in optical and {\sc Hi} wavelengths. The number of such objects is expected to be very large because the evolution of the remnants in the radiative stage takes place more slowly. Indeed, very recently first evidence of the existence of these SNRs was reported (Mavromatakis et al. \cite{mavr05}; Koo et al. \cite{koo2006}). 
  As noted above, there is large deficit of radiative radio remnants  in our Galaxy, and several optically bright SNRs are those interacting with the large-scale clouds,  and our model does not describe such remnants. The deficit of radiative SNRs in our galaxy can be explained by the irregular high-intensity background emission, which complicates the detection of extended and often fragmented and clumpy shells of SNRs.   %
   On the contrary, in the case of extragalactic observations, much more optical SNRs have been detected (Matonick \& Fesen \cite {MF}, Gordon et al. \cite{gordon}). 
 In this connection, it is important to note the very interesting finding by Duric and collaborators (Duric \cite {duric2000}; Pannuti  et al. \cite {pann2000}; Pannuti  et al. \cite {pann2002}) that very small amounts of optically-identified extragalactic SNRs are detected in radio and X-ray wavelengths; and vice versa, among the radio and X-ray selected SNRs, there are very few optically detected SNRs. %
The large number of extragalactic SNRs detected in optical wavelengths are partly the result of the fact that the sensitivity and resolution limitations severely reduce the effectiveness of radio and X-ray searches for many distant extragalactic sources. % 
At the same time, we must also take into account that many SNRs located in and near HII regions easily can be missed in optical wavelengths. Since the strong optical emission from an SNR ordinarily implies that it is in a radiative evolutionary state (H$\alpha$ is also observed   from non-radiative remnants, such as Tycho's remnant, but the emission is weak), the absence of radio emission from the optically identified extragalactic SNRs can be considered a serious argument against the mechanism of van der Laan to be the main generator of nonthermal radio emission from the radiative SNRs. It thus seems very likely that most of the optically identified extragalactic SNRs represent an assembly of old remnants in a radiative phase evolving mainly in the dense ISM that have already faded or become weaker as radio sources, so these observational data confirm the conclusions of our model. 

An opposite explanation has been suggested by Pannuti et al. (\cite {pann2000}) and Pannuti  et al. (\cite {pann2002}) that the optical remnants not possessing the radio counterpart  are those evolving in the medium with low density. If this were the case, then these remnants would have  considerably larger diameters compared with those that only radiate in the radio range. This is because the remnant should process larger volume to accumulate a column density sufficient for the onset of the radiative cooling in the low density ISM  . But the statistics of SNRs in M33 does not seem to support this possibility:  the  values of mean diameter of SNRs detected in radio and the optical SNRs without radio counterpart  are almost equal: $37.34$~pc (for 44 radio SNRs from Gordon et al. \cite{gordon}) and $38.81$~pc (for 31 optical SNRs without radio from Gordon et al. \cite{gordon98}), respectively. Unfortunately, it is difficult to take into consideration the influence of various selection effects on the statistics of extragalactic SNRs, but it is clear that many weak radio SNRs remain undetected. For example, the faintest radio SNR in the sample of SNRs in M33 has $\Sigma_{\rm 1GHz}=8.4 \times 10^{-22} \,\,\,{\rm W}\,{\rm m}^{{\rm  - 2}} \,{\rm sr}^{{\rm  - 1}} \,{\rm Hz}^{{\rm  - 1}}$, which is an order of magnitude brighter than the faintest SNR in the sample of Galactic SNRs, and from 101 Galactic shell-type and composite SNRs with known $\Sigma_{\rm 1GHz}$ $11 (11\%)$ SNRs are fainter than that remnant in M33. 
%

%XXXXXXXXXXXXXXXXXXXXXXXXXXXXXXXXXXXXXXXXXXXXXXXXXXXXXXXXXXXXXXXXXXXXXXX
\section{ Conclusions}
We have presented the model of the radio emission of shell-type SNRs evolving in the homogeneous ISM  based on the assumption that the radio-emitting electrons are accelerated by DSA mechanism  from the downstream thermal Maxwellian distribution. 
The aim of the work was not the detailed modeling of a certain SNR but to examine the capability of DSA to reproduce the general statistics of shell-type SNRs and, as far as possible, to obtain constraints on the mechanism for accelerating electrons. 
 
 1. The main properties of the model that intrinsically reflect the action of the mechanism of DSA are as follows:
\begin{itemize}
	\item SNRs evolve at nearly constant radio surface brightness  ($\Sigma \propto D^{ - (0.3 \div 0.5)}$) followed by a steep decrease (steeper than $\propto D^{ - 4.5}$). The break to a steeper power law in the $\Sigma - D$ track begins at shock Mach numbers ${\cal M}_{\rm s}\la 10$. This result is in excellent agreement with the conclusions of  Berkhuijsen (\cite{ber86}). 
 \item With increasing diameter of the remnant, the radio shell broadens, and the  center-to-limb ratio of the radio intensity over the surface of the SNR diminishes. 
 \item The radio size of the SNR with ${\cal M}_{\rm s}< 10$ is smaller than the real diameter of the shock front, and this difference increases with time.
 \item  The value of the integral spectral index of the remnant increases from the canonical value of 0.5 up to 0.6. The  values of $\alpha$ for the evolved remnants that are higher than $0.6$ can be explained in the framework of DSA in the adopted approximation,  if the remnant evolves in a preexisting cavity with a density distribution increasing radially from the center of explosion.
\end{itemize}

These features have a universal nature; i.e. they weakly depend on the input parameters of the model and practically do not depend on the poorly-known injection parameters.

2. Our model explains many properties of the empirical $\Sigma - D$ distribution for shell-type SNRs including the very large SNR candidates such as Galactic loops and N7793-S26 in the galaxy NGC 7793, which is considered as a real candidate for the hypernova radio remnant. From the shapes of the evolutionary tracks and the very wide possible range in variation of the surface brightness for a given diameter (up to 3 orders of magnitude according to Eq.~(\ref{eq39})), it follows that there is no any statistically (also, physically) meaningful $\Sigma - D$ relation that could be used as the tool for estimate the distances to SNRs.

3. Comparison of the model results with the observed $\Sigma - D$ distribution shows that most of cataloqued galactic SNRs evolve in a low-density phase of the ISM which, in turn, may imply that the interstellar space in the inner Galaxy is largely filled with a very tenuous gas as in the three-phase ISM model. If this is the case, according to the model, a population of $150-250$~pc SNRs with $ \Sigma _{{\rm 1GHz}}  \la \,\,10^{ - 22} \,\,{\rm W}\,{\rm m}^{{\rm  - 2}} \,{\rm sr}^{{\rm  - 1}} \,{\rm Hz}^{{\rm  - 1}}$ are expected to exist if the kinetic energy of the explosion is $\sim 10^{51}$~erg. However, the comparison of theoretical $\Sigma - D$ tracks with the empirical distribution shows that large fraction of observed SNRs have energies closer to $\sim 10^{50}$~erg than to $\sim 10^{51}$~erg.

 4. The model considered here predicts that, with the beginning of cooling phase, the radio emission from the remnant starts to drop rapidly, i.e.  we predict the absence of considerable radio emission from SNRs in the radiative stage of evolution. We interpreted this effect and several other observational facts  as serious evidence that the mechanism of van der Laan is not an effective generator of the radio emission of radiative SNRs. 
Observational detection of a large number of radio-quiet SNRs would serve as confirmation of the model considered in this paper.

5. By using the constructed empirical $\Sigma  - D$ distribution, we obtain an estimation of the fraction of electrons accelerated from the thermal pool in the range $(3 \div 11 )\times 10^{ - 4}$.  If acceleration takes place directly from the high energy tail of the downstream Maxwellian distribution function, then the corresponding injection momentum is estimated as $p_{\rm inj}\sim (2.7-3)\cdot p_{\rm th}$ . 

\appendix 
\section {Cox-Anderson  approximation}

The analytical approximation constructed by Cox \& Anderson (\cite {coxa82}) describes the development of an adiabatic spherical blast wave in a homogeneous ambient medium of finite pressure.  At early times, when the shock wave is strong, the structure is that of the usual Sedov self-similar solution. The law of expansion of the shock wave can be inferred from the expression ($\gamma=5/3$)
  \begin{equation}
\frac{{R_{\rm s}^3 (y_{\rm s}  - 1)^3 }}{{(3y_{\rm s}  - 2)^2 }} = \frac{{2.02E_{\rm 0} }}{{75P_{\rm 0} }} \equiv R_{\rm c}^3,
\label{A1}
\end{equation}
which describes the evolution of the post-shock pressure $P_{\rm 2}  = y_{\rm s} \cdot P_{\rm 0}$  with $R_{\rm s}$; and at large $y_{\rm s}$, it gives the Sedov solution: $P_{\rm 2}=(3\cdot2.02\, E_{\rm 0}/25)\,R_{\rm s}^{-3}$. Using the jump condition at shock front in the form $y_{\rm s}  = (5{\cal M}_{\rm s}^2  - 1)/4$, this formula can be expressed in terms of Mach number as
\begin{displaymath}
R_{\rm s}^3  = R_{\rm c}^3  \cdot \frac{{36}}{5}\frac{{\left( {{\cal M}_{\rm s}^2  - 11/15} \right)^2 }}{{\left( {{\cal M}_{\rm s}^2  - 1} \right)^3 }}.
 \end{displaymath}
For our purposes it is convenient to use this expression in another form, namely, as 
\begin{equation}
R_{\rm s}^3  = R_{\rm 0}^3  \cdot {\cal M}_{\rm 0}^2 \frac{{\left( {{\cal M}_{\rm s}^2  - 11/15} \right)^2 }}{{\left( {{\cal M}_s^2  - 1} \right)^3 }}
\label{A2}
\end{equation}
which at ${\cal M}_{\rm s} \gg 1$  takes the form of  ${\cal M}_{\rm s}  = {\cal M}_{\rm 0} (R_{\rm s} /R_{\rm 0} )^{ - 3/2}$,  where $R_{\rm 0}$ and ${\cal M}_{\rm 0}$ are the initial values of the shock radius and Mach number at the beginning of Sedov phase. %

In this approximation, the radial distribution of the density is 
\begin{eqnarray}
x(r,R_{\rm s} ) \equiv \frac{{\rho (r)}}{{\rho _0 }} = \left[ {\frac{5}{2} + \left( {x_{\rm s}  - \frac{5}{2}} \right)\,r^Q } \right]
\, r^{9/2}   \nonumber \\ \times 
\exp \left[ {\left( {\frac{{3x_{\rm s}  - 15/2}}{Q}} \right)\left( {r^Q  - 1} \right)} \right],
\label{A3}
\end{eqnarray}
where 
$ 
x_{\rm s}  = (4y_{\rm s}  + 1)/(y_{\rm s}  + 4)=4/(1 + 3/{\cal M}_{\rm s}^{\rm 2})$
 is the compression ratio at the shock front.  The history of any element of matter can be evaluated by using the equation 
\begin{equation}
r_{\rm i} (r,R_{\rm s} ) \equiv \frac{{R_{\rm i} }}{{R_{\rm s} }} = r^{5/2} \,\exp \left[ {\left( {\frac{{x_{\rm s}  - 5/2}}{Q}} \right)\,\left( {r^Q  - 1} \right)} \right],
\label{A4}
\end{equation}
which connects the original location $R_{\rm i}$ of a mass element to its present position $R=R_{\rm s}\cdot r$  when the shock radius and Mach number are $R_{\rm s}$ and ${\cal M}_{\rm s}$, respectively. In (\ref{A3}) and (\ref{A4})
\begin{eqnarray}
Q = \frac{2}{3}\frac{{x_{\rm s} (x_{\rm s}  - 1)(7x_{\rm s}  - 13)}}{{4x_{\rm s}  - 1}} \,\nonumber \\ 
=\frac{{8{\cal M}_{\rm s}^2 ({\cal M}_{\rm s}^2  - 1)(5{\cal M}_{\rm s}^2  - 13)}}{{(5{\cal M}_{\rm s}^2  - 1)({\cal M}_{\rm s}^2  + 3)^2 }}
\label{A5}
\end{eqnarray}
The expressions (\ref{A3})-(\ref{A5}) describe the structure of SNR completely. At ${\cal M}_{\rm s}  \gg 1$  they are in excellent agreement with the self-similar Sedov solution, and as noted by the authors, they were tested by the numerical calculations and are good up to $y_{\mathrm s}  \approx 2$ $({\cal M}_{\rm s}  \approx 1.4)$. 
% \end{appendix}
 The dependence of Mach number and other dynamical variables on the time can be found through the equation
\begin{equation}
\frac{t}{{t_0 }} = 1 - \frac{2}{3}{\cal M}_0^{5/3} \int\limits_{{\cal M}_{\rm 0} }^{{\cal M}_{\rm s} } {\frac{{\left( {z^2  - 1/5} \right)dz}}{{\left( {z^2  - 1} \right)^2  \cdot \left( {z^2  - 11/15} \right)^{1/3} }}}, 
\label{A6}
\end{equation}
 which follows from Eq.~(\ref{A2}) and the definition $c_{\rm s} \,dt = (1/{\cal M}_{\rm s})\,(dR_{\rm s}/d{\cal M}_{\rm s})\,d{\cal M}_{\rm s}$.
 It is important to note that at high values of ${\cal M}_{\rm s}$ this equation corresponds to dependence ${\cal M}_{\rm s}  = {\cal M}_0  \, [2.5 \,(t/t_0) -1.5]^{ - 3/5}$, but not $ {\cal M}_{\rm s} = {\cal M}_0 \,( t/t_0 )^{ - 3/5} $  as in case of a pure Sedov law of expansion.  Utilization of such a form for the dependence of Mach number on time corresponds to using the offset power-law formalism (Cioffi et al.  \cite {cioffi}), which describes the very beginning of the smooth transition from the phase of free expansion to Sedov phase. As a result, we can assign low values to the parameter $\chi$ that allows us to cover a large part of the life of the remnant.

\begin{acknowledgements} 
The author thanks Ms. Maya Omarova for her help in preparation of the manuscript, Dr. Elchin Jafarov for the help with the data collection, and Dr. Jurii Turovskii for his help with LaTex.  The author also expresses his gratitude to Prof. John Dickel, the referee of this paper, for his detailed comments and suggestions, which greatly improved this paper.
\end{acknowledgements}

%%%%%%%%%%%%%%%%%%%%%%%%%%%\newpage
%\end{document}

%%%%%%%%%%%%%%%%%%%%
\section {List of Symbols}
%\begin{table}
%\caption{}
%\label{tab3}
%\medskip
%\centering
%\begin{minipage}{8cm}

%%%%%%\begin{tabular}{ll}
$B_{0}$  - interstellar magnetic field.\\
$B_{ -5}$ - interstellar magnetic field in units of $10^{-5}$ G.\\
$C $ - parameter describing adaiabatic cooling of accelerated electrons [Eq.~(\ref{eq12})].\\ 
$c_{\rm 1}  =\bar \mu\, \psi ^2 {\rm v}_{09}^2  \cdot B_{{\rm - 5}} /\nu _9$ - parameter of the problem [below Eq.~(\ref{eq22})].\\  
$D_{s}$ - SNR diameter.\\
$D_{break}$ - SNR diameter at which break in the $\Sigma - D$ track occurs [Eqs.~(\ref{eq37}), (\ref{eq38})].\\
$d_{kpc}$ - the distace to SNR in kpc.\\
$f_{a} $ - distribution function of accelerated by DSA electrons.\\
$f $ - radially dependent distribution function [Eq.~(\ref{eq14})].\\
$f_{m} $ - Maxwellian distribution function of electrons.\\
$H $ - Heaviside step function.\\
$E_{\rm SN}$ - initial kinetic energy of the SN shell.\\
$E_{\rm 051}$ - initial kinetic energy of the SN shell in $10^{51}$ erg.\\
$k,k_{\rm 0}$ - power law index in the dependence of the magnetic field on the density $B\propto \rho^{k}$.\\
$K_{e} $ - coefficient of the power-law spectrum [Eq. (\ref{eq16})].\\ 
$l_{\nu}$ - undimensional radio flux density [Eqs. (\ref{eq23}), (\ref{eq25})].\\
$\cal M, M_{\rm s} $ - shock wave Mach number.\\
${\cal M}_{cool} $ - Mach number at which thin shell formation begins [Eq.~(\ref{eq35})]. \\
${\cal M}_{max} $ - Mach number at which $l_{\nu}$ reaches its maximum.\\
${\cal M}_{0} $ - initial shock wave Mach number.\\
$M_{ej} $ - ejected mass at SN explosion.\\
$m_{ej} $ - ejected mass at SN explosion in units of solar mass.\\
$n_{e} $ - electron number density.\\
$n_{e0} $ - electron number density in the ISM.\\
$P_{0}$ - total pressure of the ISM [Eq. (\ref{eq29})].\\
$P_{04}$  - total pressure $P_{0}$ in units of $10^{4}$\, $\rm K \,cm^{-3}$.\\
$P_{th} $ - thermal gas + cosmic rays pressure.\\
$P_{M} $ - magnetic field pressure.\\
$P_{2} $ - post-shock pressure.\\
$p_{inj} $ - injection momentum of electrons.\\
$p_{max} $ - maximum momentum of accelerated electrons.\\
$p_{th} $ - thermal momentum of electrons.\\
$p_{i0} $ - injection momentum when shock radius was $R_{0}$.\\
$Q$ - parameter of the Cox-Anderson approximation [Eq. (\ref {A5})].\\
$q $ - spectral index of the momentum distribution function of accelerated by DSA electrons [Eq. (\ref{eq1})].\\
$q_{i} $ - spectral index of the momentum distribution function of electrons accelerated when shock radius was $R_{i}$.\\
$R$ - radial coordinate from center of the SNR.\\ 
$R_{c}$ - characteristic radius of SNR [Eq. (\ref {A1})].\\
$R_{s}$ - SNR radius.\\ 
$R_{0}$ - initial radius of SNR at which Sedov phase begins.\\
$R_{i}$ - original radial position of mass element.\\
$R_{in}$ - present radius of mass element which was first shocked at $R_{0}$.\\
$r_{in}$ = $R_{in}/R_{s}$.\\ 
$r_{i}$ - normalized original location of mass element $=R_{i}/R_{s}$ [Eq. (\ref {A4})]\\
$r$ - normalized radial coordinate from center of the SNR: $r=R/R_{s}$.\\
$S_{\nu}$ - radio flux density [Eq. (\ref{eq23})].\\
${\rm v}_{{\rm 0s}}$ - initial shock velocity.\\ 
${\rm v}_{{\rm 09}}$ - initial shock velocity in the units of $10^{9}$ cm/s.\\
$x$ - normalized radially dependent density distribution [Eq.~ (\ref {A3})].\\
$x_{s}$ - shock compression ratio ([Eq. \ref {eq3}] and Section 3.1).\\ 
$\alpha$ - spectral index of the radio emission, determined as $S_{\nu }\propto\nu^{-\alpha}$.\\
$\beta$ - ratio of thermal gas + cosmic ray pressures to the magnetic field pressure(Sect. 3.1).\\
$\varepsilon _\nu,  \varepsilon _{{\rm \nu 0}}, \varepsilon_\nu (r)$ - radio synchrotron emission coefficient [Eqs.~(\ref{eq18}), (\ref{eq20}-\ref{eq22})].\\
$\varepsilon$   $= p_{{\rm inj}} /p_{{\rm max}}$ [Eq.~(\ref{eq7})].\\
$\eta$ - injection parameter [Eq.~(\ref{eq5})].\\
$\kappa$ - spatial diffusion coefficient.\\ 
$\bar \mu$ - mean mass per particle in units of the proton mass.\\
$\mu _{\rm e}$ - mean mass per electron in units of the proton mass.\\
$\Sigma$ - radio surface brightness of the SNR.\\ 
$\Sigma _{0\nu }$ - radio surface brightness of the SNR [Eq.~(\ref{eq27})].\\ 
$ \Sigma _{{\rm 1GHz}}$ - radio surface brightness of the SNR at $1$ GHz.\\ 
$\sigma _{\nu}$ - undimensional surface brightness [Eqs.(\ref{eq26}), (\ref{eq28})].\\
$\chi$ - parameter characterizing the transition from free expansion to the Sedov phase.\\
$\psi$ - injection parameter [Eq.(\ref{eq6})].\\ 

%\end{tabular}
%\end{minipage}
%\end{table}
%%%%%%\end{tabular}
%The subscript $i$  denotes the values at the original location $R_{i}$ but now found at $R=r\,R_{s}$ inside the SNR, the subscript $0$  denotes the initial values at the beginning of the Sedov phase at the radius $R_{0}$ with Mach number ${\cal M}_{0}$. 
%\end{minipage}

\end{document}